\documentclass[twocolumn,trackchanges]{aastex63}

\usepackage{graphicx}
\usepackage{amsmath}
\usepackage{psfig}
\usepackage{color}
\usepackage{icomma}

\def\nsec{\textrm{ns}}
\def\usec{\mu\textrm{s}}
\def\msec{\textrm{ms}}
\def\sec{\textrm{s}}
\def\yr{\textrm{yr}}
\def\Megasec{\textrm{Ms}}
\def\kyr{\textrm{kyr}}
\def\Myr{\textrm{Myr}}
\def\Gyr{\textrm{Gyr}}

\def\um{\mu\textrm{m}}

\def\cm{\textrm{cm}}
\def\meter{\textrm{m}}
\def\km{\textrm{km}}

\def\kpc{\textrm{kpc}}
\def\Mpc{\textrm{Mpc}}

\def\kms{\textrm{km\ s}^{-1}}

\def\gram{\textrm{g}}

\def\sr{\textrm{sr}}

\def\Kelv{\textrm{K}}

\def\Msun{\textrm{M}_{\odot}}

\def\erg{\textrm{erg}}
\def\keV{\textrm{keV}}
\def\gcm2{\textrm{g}\,\textrm{cm}^{-2}}
\def\rhoUnits{\textrm{g}\,\textrm{cm}^{-3}}
\def\Lsun{\textrm{L}_{\odot}}

\def\AU{\textrm{AU}}

\def\mas{\textrm{mas}}

\def\Rsun{R_{\odot}}

\def\Rstar{R_{\rm em}}
\def\Rlens{R_L}
\def\Tlens{T_L}
\def\Roccult{R_{\rm occult}}

\def\xStar{x_{\rm em}}
\def\Lstar{L_{\rm em}}
\def\Mstar{M_{\star}}
\def\Fstar{F_{\rm em}}
\def\Tstar{T_{\rm em}}

\def\LBL{L_{\rm BL}}
\def\TBL{T_{\rm BL}}
\def\LDisk{L_{\rm disk}^{\rm app}}
\def\Tdiskin{T_{\rm disk}^{\rm in}}

\def\tflare{t_{\rm flare}}

\def\tshadow{t_{\rm shadow}}
\def\twindow{t_{\rm window}}
\def\twindowflare{t_{\rm window}^{\rm flare}}
\def\twindowshadow{t_{\rm window}^{\rm shadow}}
\def\tres{t_{\rm res}}

\def\Fobs{F_{\rm obs}}

\def\thetashadow{\theta_{\rm shadow}}
\def\thetalens{\thetashadow}
\def\thetaimage{\theta_{\rm image}}
\def\thetastar{\theta_{\rm em}}
\def\thetainf{\theta_{\infty}}

\def\vrel{v_{\rm rel}}
\def\vrand{v_{\rm rand}}
\def\vlens{v_L}
\def\alens{a_L}
\def\Nlens{N_L}
\def\nlens{n_L}
\def\Plens{P_L}
\def\Sigmalens{\Sigma_L}

\def\Mlens{M_L}
\def\Mswarm{M_{\rm swarm}}

\def\aNS{a_{\rm NS}}
\def\abin{a_{\rm bin}}
\def\Pstar{P_{\rm em}}
\def\Pbin{P_{\rm bin}}

\def\Mdonor{M_{\rm donor}}
\def\MNS{M_{\rm NS}}
\def\Mbin{M_{\rm bin}}

\def\Fluenceshadow{{\cal F}_{\rm shadow}}
\def\Fluenceflare{{\cal F}_{\rm flare}}
\def\Fluenceexcess{{\cal F}_{\rm excess}}

\def\Fluxflare{F_{\rm flare}}
\def\Fluxshadow{F_{\rm shadow}}

\def\Aeff{A_{\rm eff}}

\def\NBL{N_{\rm BL}}
\def\NBLnorm{\overline{N_{\rm BL}}}
\def\Ndisk{N_{\rm disk}}
\def\Nback{N_{\rm back}^{\rm eff}}
\def\SNR{(S/N)}

\def\Gammaflare{\Gamma_{\rm flare}}
\def\Gammacoll{\Gamma_{\rm col}}
\def\Pflare{{\cal P}_{\rm flare}}

\def\AreaFactor{{\cal A}}

\def\zetaabs{\zeta_a}
\def\zetageom{\zeta_g}

\def\tdrift{t_{\rm drift}}

\def\ga{\gtrsim}
\def\la{\lesssim}

\def\endash{\text{--}}

\def\edit1{}
\newcommand{\editTwo}[1]{#1}
\newcommand{\editThree}[1]{#1}

\shorttitle{Lens Flare}
\shortauthors{Lacki}

\begin{document}

\title{Lens Flare: Magnified X-Ray Binaries as Passive Beacons in SETI}
\author{Brian C. Lacki}
\affiliation{Breakthrough Listen, Astronomy Department, University of California, Berkeley, CA, USA}
\email{astrobrianlacki@gmail.com}

\begin{abstract}
Low mass X-ray binaries (LMXBs) containing neutron stars are both extremely luminous and compact, emitting up to $\sim 10^6\ \Lsun$ within a kilometer-scale boundary layer.  This combination allows for easy modulation, motivating X-ray SETI.  \editThree{When X-ray lenses with radii $100 \endash 1,000\ \km$} magnify the LMXB boundary layer, it brightens by a factor of several thousand for \editThree{a fraction of a} second.  In addition, there should be occultation events where the neutron star is blocked out.  Passive X-ray lenses could require little \edit1{internal power} and the LMXB light source itself shines for millions of years, \edit1{with potential for an effective beacon for interstellar communication}.  A very large number of lenses would be needed to ensure \editThree{frequent signals in} all directions, however, and gathering material to construct them could be very difficult.  Avoiding collisions between lenses\edit1{,} aiming them\edit1{, and building and maintaining their precise shapes} pose additional challenges.  \editThree{``Lens flares''} of \editThree{bright} LMXBs are easily detectable in the Galaxy, although they would be rare events, occurring \editThree{perhaps} once per decade.  Our \editThree{sensitive} X-ray instruments could detect the \editThree{eclipses of Galactic LMXBs and possibly intergalactic flares}, but it is unlikely they would be observing the LMXB \editThree{at the right time}.
\end{abstract}


\section{Introduction}
Now and again, conducting the Search for Extraterrestrial Intelligences (SETI) in X-rays is proposed \citep{Corbet97,Carstairs02,Hippke17-XSearch}.  Extraterrestrial intelligences (ETIs) may have several reasons to prefer X-rays.  With very small wavelengths, X-rays can be beamed very effectively by a diffraction-limited instrument \citep[c.f.,][]{Skinner01}, an advantage that limits losses by diffraction and maximizes information transmission per unit energy \citep{Hippke17-OptimalNu,Hippke17-XSearch}.  They also allow more distinct temporal modes (with time resolutions down to attoseconds), another advantage when maximizing information transmission rates \citep[c.f.,][]{Caves94,Hippke17-OptimalNu,Hippke18-MaxInfo}.  But the most common reason for proposing X-ray SETI is that they are the main form of luminosity of compact objects, particularly neutron stars (NSs).  Even a small amount of matter dropped on a NS can produce a bright signal, luminous enough to be seen across the Galaxy \citep{Corbet97}.  Of course, X-ray SETI is always a search for societies with capabilities far beyond our own: we do not have large diffraction-limited X-ray optics, attosecond time resolution on X-ray detectors, and certainly not a spare neutron star lying around.

ETIs do not need complicated transmitters that generate tremendous amounts of power to construct a ``beacon'' to grab the attention of their neighbors.  Rather than directly generate radiation, ETIs can instead construct passive beacons.  These modulate an extant luminous source using a large solid structure requiring no power \edit1{and with reduced need for moving parts}.  \citet{Arnold05} presented one of the first passive transmitters concepts: a planet-sized megastructure with an unusual shape, detectable by its unusual light curve when it passes in front of its sun.  The modulation of bright neutron stars by transiting objects is a passive beacon much more luminous than the Sun \citep{Chennamangalam15,Imara18}.  Another simple example is a heliograph, a mirror that reflects sunlight to a target \citep{Lacki19-Glint}.  As passive beacons, they need \edit1{less} maintenance to remain capable of broadcasting signals.  The only threats are space weathering\edit1{, structural deformation for optical elements,} and possibly guidance to ensure a stable aim and collision avoidance.  In contrast, a beamforming or laser system might break down over millennia.  As Drake's equation has taught from the beginning of modern SETI, only long lasting technosignatures are likely to be detected \citep[e.g.,][]{Bracewell60,Sagan73,Forgan11}.

Linear optical systems like mirrors and lenses can at best only preserve an object's surface brightness, a fundamental limitation from thermodynamics.  Fully modulating a source's luminosity with passive optics requires a structure as big as the emitting source.  It is even possible to greatly boost a source's luminosity temporarily by using a lens with proportionally greater area than its source.  The resulting transients when a lens passes in front of its host is called here a ``lens flare'', and it is the premise of this paper.  But lens flares from stars require megastructures much larger than a sun.

X-ray emitting sources are highly advantageous for these passive beacons, because of the Stefan-Boltzmann law.  The accreting NSs in Low Mass X-ray Binaries (LMXBs) can exceed $10^5\ \Lsun$ in luminosity, mostly in X-rays.  While some of the luminosity comes from a large accretion disk, more than half the luminosity of LMXBs with NSs can arise from a boundary layer (BL) where accreting matter settles onto the NS \citep{Sunyaev86}.  The boundary layer may take the form of an equatorial belt of width a few kilometers within the accretion disk \citep{Popham01}, or spreading layers in mid-latitude belts from matter torqued away from the accretion disk \citep{Inogamov99}.  Thus, LMXBs are attractive not just because they are bright, as noted by \citet{Corbet97}, but because they are small too.  A lens as big as a small city could double its already high luminosity.  If lenses as big as the planet-sized screens of \citet{Arnold05} can be built, the resultant lens flares could \editThree{appear} brighter than an entire galaxy.\footnote{Brightness temperatures exceeding $10^{35}\ \Kelv$ are achieved in radio transients \citep[e.g.,][]{Lorimer07}.  The nanoshot radio emission of pulsars like the Crab may come from meter-scale regions that glow for a few nanosecond, or somewhat larger regions that are relativistically beamed \citep{Hankins03,Hankins07}.  Perhaps a kilometer-scale lens could boost their emission to much higher levels still.  The trouble is that these sources are transient.  Not only is the radio emission off most of the time, but the location of the meter-scale emission is probably a random spot in a much larger volume.  Unless the emission regions are stable, a lens would almost certainly magnify empty space instead.  If stable emission regions do exist and are lensed, however, the resulting lens flare would appear superficially like a Fast Radio Burst.}  Some galaxies have ultraluminous X-ray sources (ULXs) with effective luminosities of $10^{39} \endash 10^{40}\ \erg\ \sec^{-1}$ or more.  Many of these are now thought to also be accreting NSs that beam most of their X-ray emission towards us \citep[e.g.,][]{Begelman06,Bachetti14,Israel17}.  ULXs might illuminate even more powerful beacons.

I propose that ETIs could exploit the huge surface brightness of the boundary layer by placing a swarm of giant X-ray lenses around a\editTwo{n} LMXB.\footnote{\editThree{The structures may actually be arrays of smaller elements, whether refractive or diffractive, rather than single lenses hundreds of km in radius.  I nonetheless refer to them as ``lenses'' because their function would be similar.}}  Each lens serves as a collimator, with a tight beam of hard X-rays as output.  As the lenses orbit the LMXB, observers would see an X-ray transient as the lens occults the boundary layer and the collimated beam sweeps over them.  A numerous enough swarm of lenses can ensure that observers viewing from any angle will be able to see a flare.  The initial construction of a lens swarm would be difficult, but the flares would be easily visible anywhere in the Galaxy, and possibly much further beyond.  \edit1{The luminosity source for the lens swarm is also very long-lived, although the lifetime of the beacon depends on the lifetime of the swarm without performance degradation.} The X-ray luminosity has a lifespan thought to be from a few Myr for ultracompact LMXBs \citep{Bildsten04} to $\sim 1\ \Gyr$ (though possibly with a duty cycle of $\sim 1\%$; \citealt{Webbink83,Pfahl03}), and a maximum lifetime of 100 Myr for LMXBs with luminosities $10^{38}\ \erg\,\sec^{-1}$ \citep{Gilfanov04}.

Building a lens swarm requires advanced technology.  Long-distance interstellar travel is necessary to reach the LMXB in the first place.  \editTwo{If ETIs exist in globular clusters with LMXBs \citep[c.f.,][]{diStefano16}, the distances would be only of order a parsec.} In addition, the ETIs must be capable of building efficient and large optical elements for hard X-rays.  This may be achievable by using a Fresnel zone plate -- essentially a membrane with a specific series of alternately absorbing and transparent concentric rings -- or a similar structure as a lens.  Zone plates are proposed as an enabling technology for large, lightweight X-ray telescopes \citep{Skinner01,Skinner02,Skinner10}.  The main disadvantage of this simple technology is chromatic aberration, although it is possible to counteract it over a limited energy range using a refractive element \citep{Skinner02,Wang03}.  They also have long focal lengths, but this is an asset because only lenses far from the LMXB can survive the onslaught of LMXB radiation.  I will consider achromatic lenses\editThree{, simple lenses with chromatic aberration,} and uncorrected zone plates.

The next section describes the theory of lens swarms: the optical configuration, orbits of the lenses, and the problems of building and maintaining the swarm.  Section~\ref{sec:ObservingLensFlares} explores our capabilities of observing lens flares with past, current, and near-future X-ray instruments.  Finally, the conclusion (Section~\ref{sec:Conclusion}) raises the question of trade-offs in building a lens flare as opposed to a simpler but much harder-to-detect structure like a stellar occulter.

\section{The lens system}
\subsection{Basic optical considerations for an achromatic lens}
\begin{figure}
\centerline{\includegraphics[width=8cm]{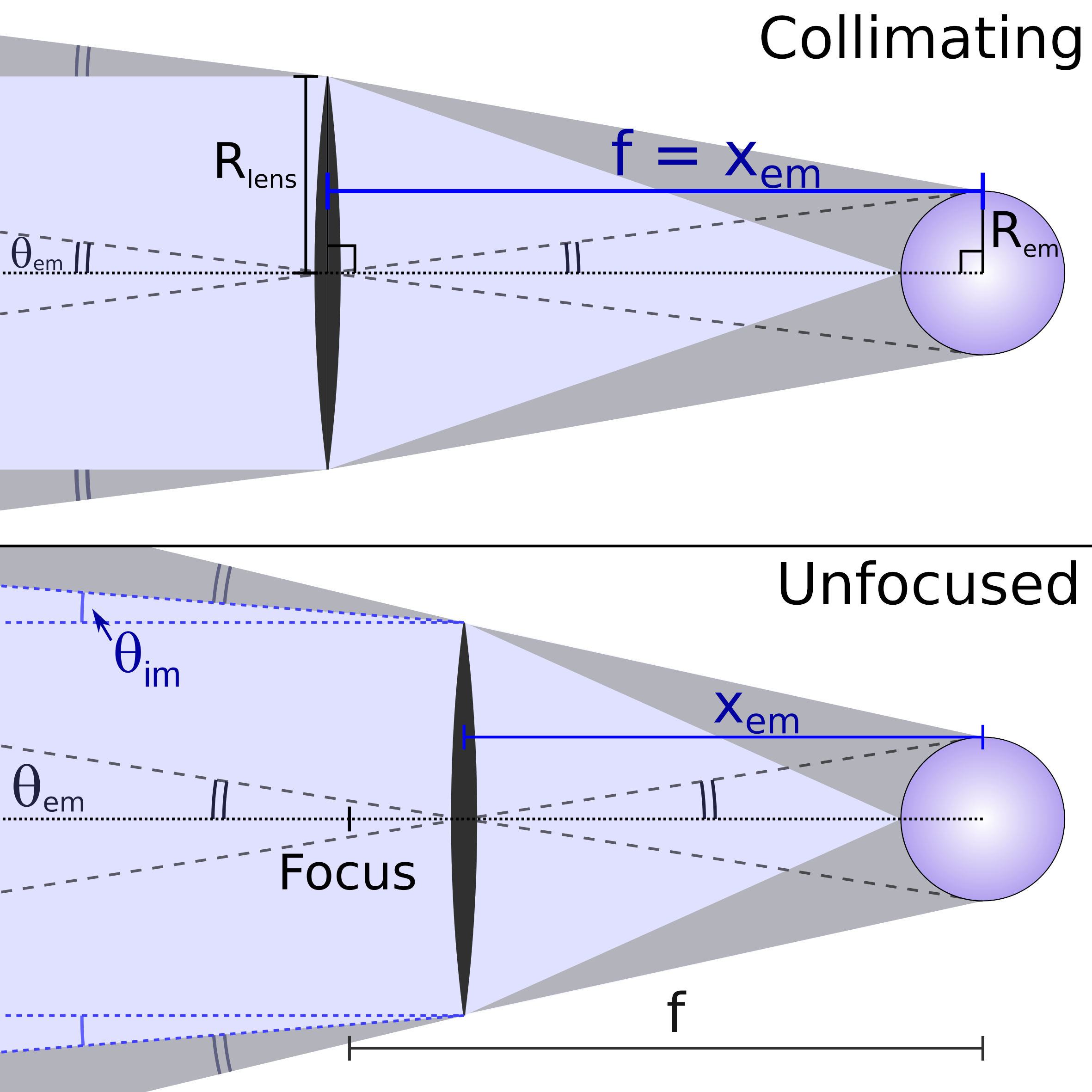}}
\figcaption{Geometry of the lens system.  At top, the source is at the focus of the lens, and the lens acts as a collimator; whereas, at bottom, the light from each point on the unfocused source (as in the light blue regions) diverges at infinity within a cone of opening angle $\thetastar$.\label{fig:SideGeometry}}
\end{figure}

\begin{figure*}
\centerline{\includegraphics[width=6cm]{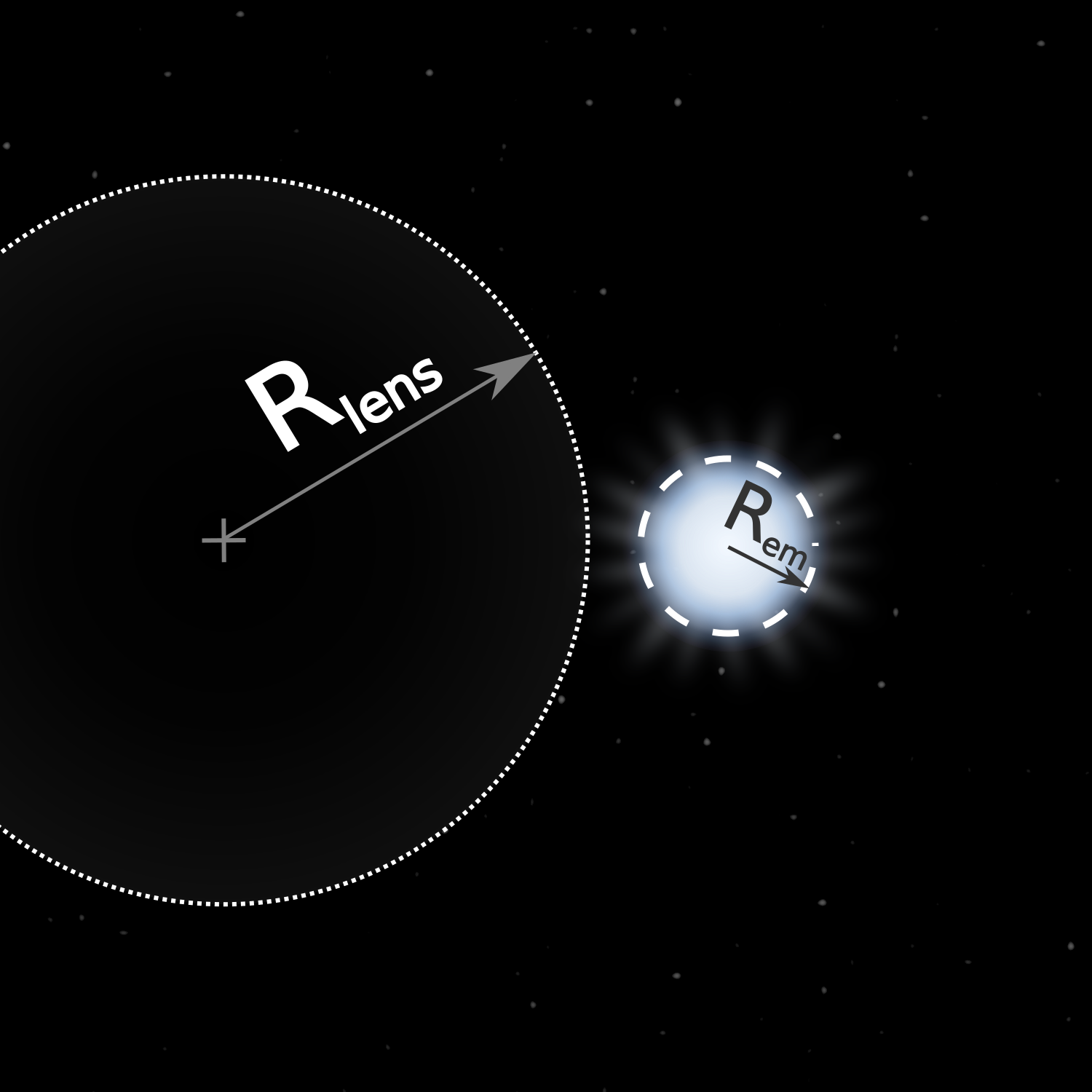}\,\includegraphics[width=6cm]{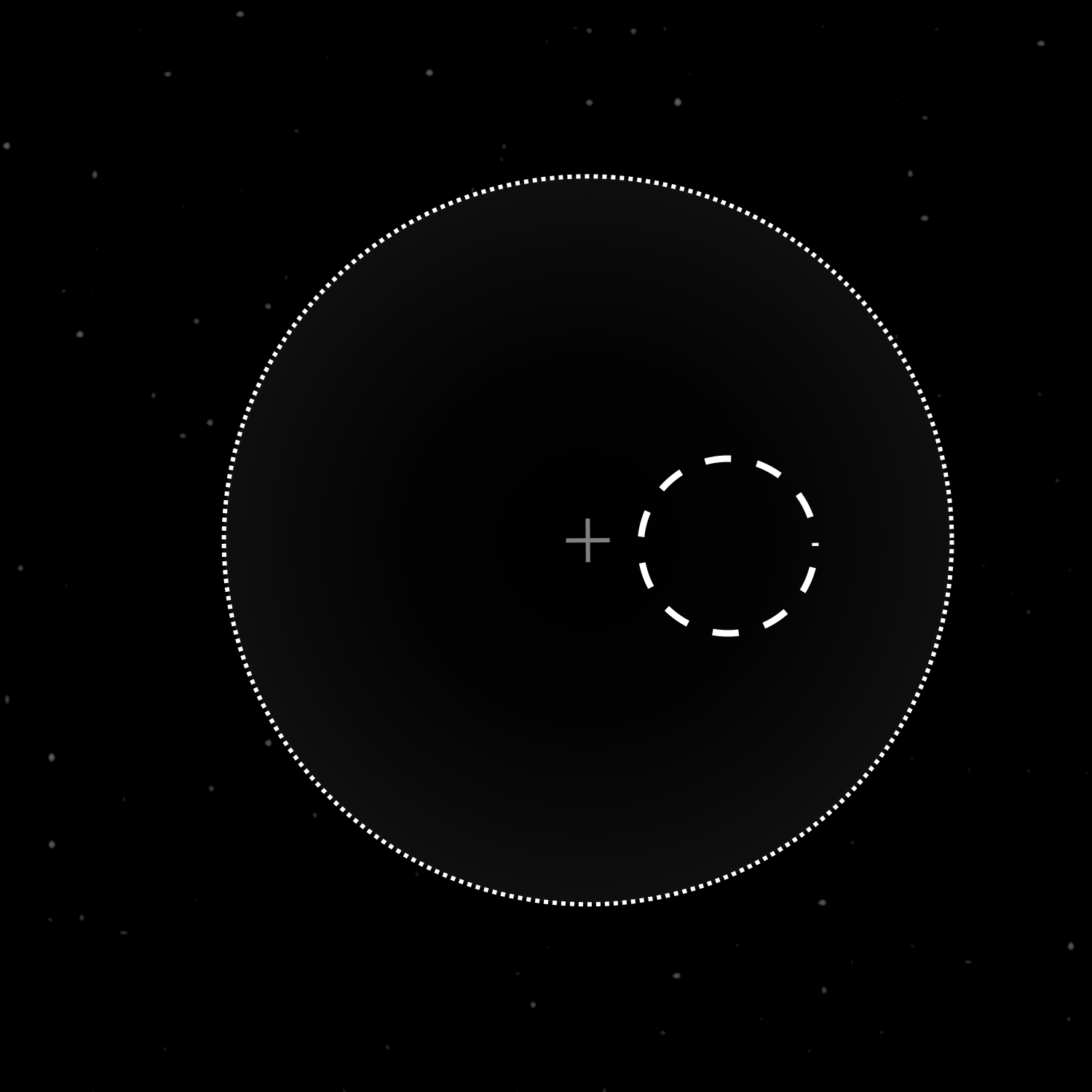}\,\includegraphics[width=6cm]{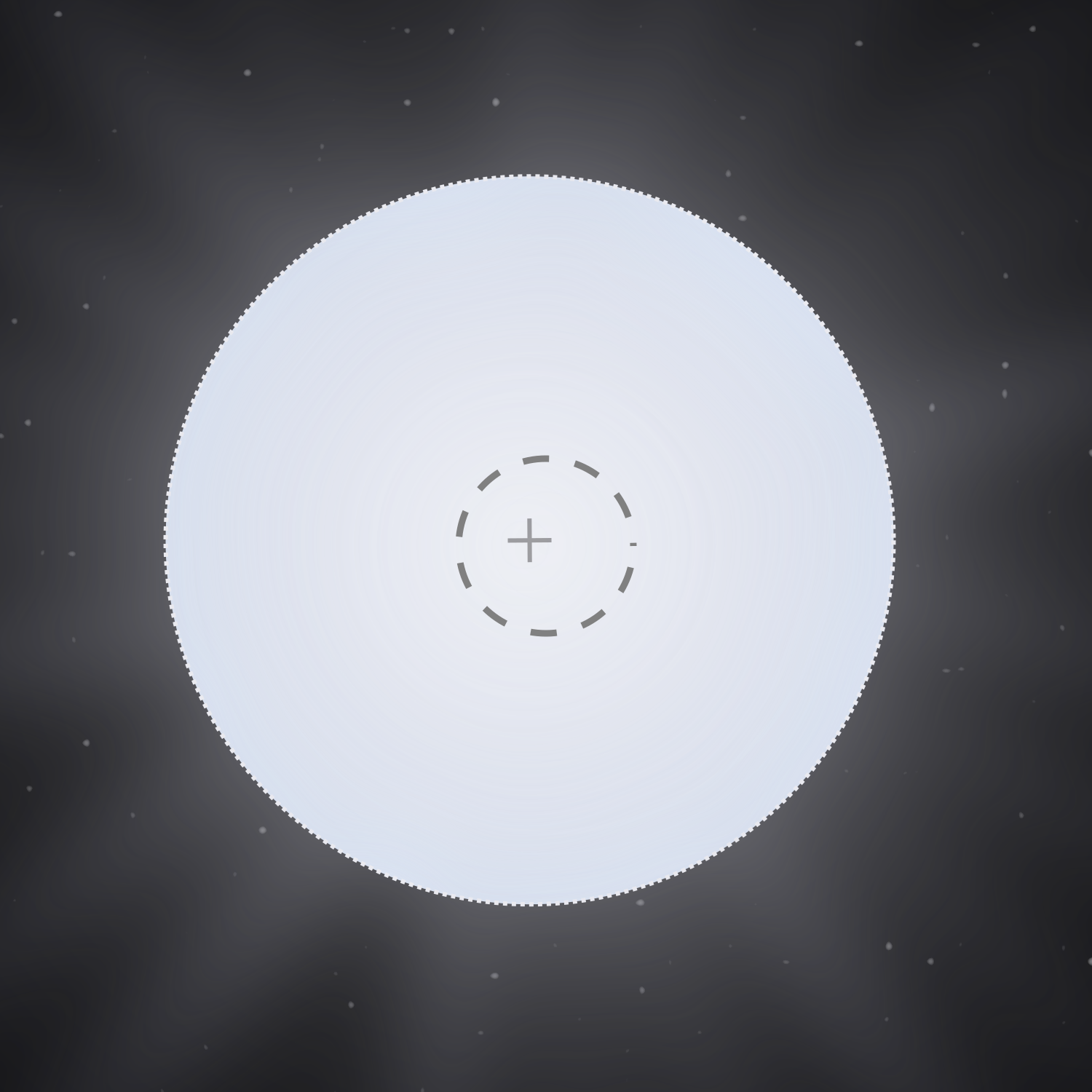}}
\figcaption{Face-on view of an achromatic lens and its effects on the source.  If the source is at the focus, then the lens has the same surface brightness as whatever is directly behind it (grey cross).  The source appears to have its normal brightness if the lens does not occult the source (left), and is eclipsed if the lens occults it but magnifies empty space (center).  When the lens is centered over the source, the apparent luminosity of the system is enhanced greatly as the entire lens appears to have the source's surface brightness(right).\label{fig:FaceOn}}
\end{figure*}

The X-ray beacon can be a very simple optical system.  Suppose a large, non-absorbing lens with a very large focal length $f$ is placed at distance $\xStar$ in front of a central \editThree{illuminating} source, in this case the NS in an LMXB (Figure~\ref{fig:SideGeometry}, top).  The thin lens equation gives the distance $x$ in front of the lens at which \editThree{a magnifying} image is formed: $1/f - 1/\xStar = 1/x$.   Now, if the lens is placed one focal length from the LMXB ($\xStar = f$), the image will be formed at infinity, with infinite magnification.  What this means is that all the light captured by the lens from one point on the LMXB's surface gets sent in one direction on the distant sky.  The lens has become a collimator for the LMXB's light.  From the observer's point of view, the entire lens appears to light up with the same surface brightness as whatever is directly behind its center (Figure~\ref{fig:FaceOn}, right).  If a circular lens has a radius $\Rlens$ is viewed when it is directly in front of a spherical emitting region of radius $\Rstar$, the luminosity will appear to increase by a factor of $\AreaFactor \equiv (\Rlens/\Rstar)^2$, the ratio of the projected areas of the lens and the emitting region.

Since we cannot resolve the system, \editThree{a NS passing directly behind a large lens ($\Rlens \gg \Rstar$) appears} as an eclipse, with an extremely bright flare in the middle.  First, when the \editThree{NS passes behind the edge of the lens}, we lose the light from the NS (Figure~\ref{fig:FaceOn}, middle).  Because the center of the lens is in front of empty space, we see the lack of light from that empty space instead.  We have entered the shadow of the lens, which has angular radius $\thetashadow = \Rlens / \xStar$.  When the center of the lens occults the emitting region, the lens suddenly brightens to an apparent luminosity that is $\eta \AreaFactor$ times brighter than the magnified source, where $\eta$ is the \editThree{efficiency} of the lens.  This corresponds to the image of the LMXB boundary layer, which \editTwo{is beamed into a cone with angular radius} $\thetaimage = \Rstar / \xStar$.  Then, the center of the lens is in front of empty space again, even though the emitting region is still behind the lens periphery, and we are in eclipse again.  Finally, the LMXB returns to the normal brightness once the lens is completely past the NS.  Note that the boundary layer only emits a fraction of the bolometric flux from the LMXB, which can be as high as $\sim 70\%$ or as low as a few percent, but frequently is of order $50\%$ \citep{Sunyaev86,Revnivtsev06}.  The rest, from the larger accretion disk, will remain unaffected by the transit.  The boundary layer is expected to dominate the flux in harder X-rays of a few keV, though.

A lens trades the flare brightness for the fraction of the sky where the flare is visible, ensuring conservation of energy.  If the illumination source has a flux of $\Fstar$ \editThree{and has a relative velocity $\vrel$ with the lens}, the lens shadow creates a fluence deficit of:
\begin{equation}
|\Delta \Fluenceshadow| = \frac{\Rlens}{\vrel} \Fstar. 
\end{equation}
If we \editThree{are fortunate}, though, and the center of the lens has an impact parameter $< \Rstar$, we observe a flare with fluence
\begin{equation}
\Delta \Fluenceflare = \frac{\Rstar}{\vrel} \Fstar \times \left(\frac{\Rlens}{\Rstar}\right)^2 = \left(\frac{\Rlens}{\Rstar}\right) |\Delta \Fluenceshadow|.  
\end{equation}
The flare \editThree{has} the much larger effect \editThree{on fluence during its occultation}.  For most sightlines where the lens occults the NS, the lens center misses the NS, and only the short eclipse with no flares is observed.  The probability of an eclipse without a flare is $1 - \Rstar / \Rlens$.  This compensates for the brightness of the flare, so the average brightness of an LMXB transited by a dense, isotropically distributed swarm of orbiting \editThree{efficient} lenses is the same as an unlensed LMXB.

\subsection{Lenses with chromatic aberration}
\editThree{The X-ray index of refraction of atomic matter deviates by a small difference $\delta$ from $1$.  Although $\delta$ is small, it may be sufficient to build a simple lens because of the huge focal length, although the element would need to be a thin Fresnel lens to avoid X-ray absorption.  Since $\delta$ is highly dependent on photon energy $E$, however, a single lens will suffer severe chromatic aberration.  Typically, the dependence of $\delta$ is approximately $\propto E^{-2}$ \citep[e.g.,][]{Braig12}.  As a result, the focal length is $f(E) \approx \bar{f} (E / \bar{E})^2$,} where $\bar{f}$ is some reference focal length for reference energy $\bar{E}$.  I adopt the convention $\bar{f} = \xStar$, so that the lens acts as a collimator for $\bar{E}$.

From the thin lens equation, the image is formed at a distance $x$ from the lens, where $x$ can be positive or negative.  The light incident on the lens either converges at $x$ and then diverges beyond it, or it diverges from the virtual image at $-|x|$.  Either way, the light from a point source ultimately forms a cone extending to infinity with opening angle:
\begin{equation}
\thetaimage \approx \frac{\Rlens |\xStar - f|}{f \xStar},
\end{equation}
assuming that \editThree{$\xStar \gg \Rlens |\bar{E}^2 - E^2|/E^2$}, which is true except at very low (optical) energies.  In addition, because the emitting source is not a point, this cone is broadened by an angle $\thetastar = \Rstar / \xStar$, which contains all sightlines that pass from the emitting region through the center of the lens.  Thus, if the emitting region has uniform brightness, the light intercepted by the lens diverges from the image as a cone with opening angle $\thetainf \equiv \thetaimage + \thetastar$ (Figure~\ref{fig:SideGeometry}, bottom).\footnote{\editTwo{The apparent luminosity is not uniform within the cone, as the observer sees light from an aperture of angular width $\thetaimage$, which can include empty space.  This effect is not treated, and it is relatively minor most of the time unless $\thetaimage \sim \thetastar$.}}  For \editThree{an uncorrected lens}, this angle varies as energy as:
\begin{equation}
\thetainf \approx \frac{\Rstar}{\xStar} \left(1 + \frac{\Rlens}{\Rstar} \editTwo{\frac{|\bar{E}^2 - E^2|}{E^2}}\right).
\end{equation}
For photon energies very near $\bar{E}$, $\thetainf \approx \thetastar$ as in the achromatic case.  But away from that energy, the collected light diverges into a wider cone.  If the lens is large ($\Rlens \gg \Rstar$), photons with $E \gg \bar{E}$ diverge into an angle $\sim \thetalens$.  The equality $\thetainf = \thetalens$ holds for $E = \bar{E} \sqrt{\Rlens/\Rstar}$ and also $\bar{E} / \sqrt{2 - \Rstar/\Rlens}$, for which the lens essentially has no effect on a uniformly bright source.  Furthermore, the light actually diverges into an angle much larger than $\thetalens$ at low energies, with $\thetainf \gg \thetalens$ when $E \ll \bar{E}/2$.  In these energy ranges, the lens appears to glow weakly even if it is not covering any part of the emitting region.

\begin{figure*}
\centerline{\includegraphics[width=18cm]{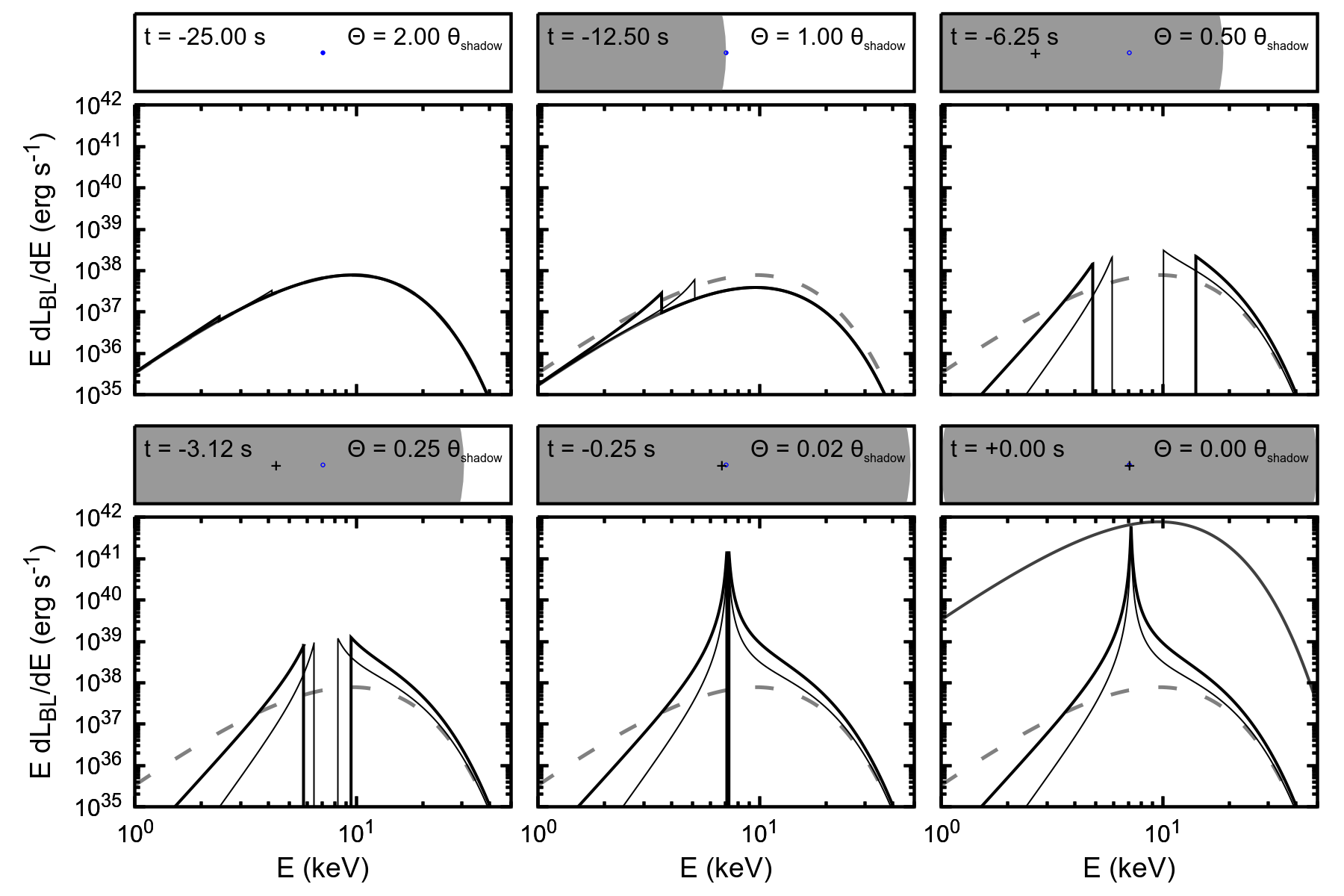}}
\figcaption{Real-time evolution of boundary layer flux spectrum.  The natural emission is a Wien spectrum with $k_B \Tstar = 2.4\ \keV$ and $\Lstar = 10^{38}\ \erg\,\sec^{-1}$.  \editThree{A source with relative speed $\editTwo{80}\ \kms$ passes directly behind} \editThree{a} lens with $\Rlens/\Rstar = 100$ and $\eta = 1$ .  The \editThree{thick} black line is the spectrum of a \editThree{diffractive} zone plate transit with $\bar{E} = 3 k_B \Tstar$, \editThree{the thin black line is the spectrum of a single refractive lens with chromatic aberration,}  the dark grey solid line is the spectrum of an achromatic lens transit, and the dashed grey line is the unmodified BL spectrum.  When the NS is occulted, the \editThree{chromatic element spectra are} filled in from high and low energies before eroding again, whereas the achromatic lens spectrum disappears at all energies except for the brief flare when all energies are greatly magnified.  A subplot on top shows the relative positions of the lens (grey disk, centered on the black cross) compared to the NS (small blue disk).
(An animation of this figure is available.)\label{fig:SpectrumMovie}}
\end{figure*}

Let $d\Lstar/dE$ be the luminosity spectrum of the lensed region, and let $\Theta$ be the angle between the source, the lens center, and the \editTwo{sightline through the lens for an} observer at infinity.  \editThree{The lens intercepts a \editTwo{specific} luminosity of $d\Lstar/dE \times (\pi \Rlens^2) / (4 \pi \xStar^2)$, which is then emitted into a cone covering $\pi \thetainf^2$ steradians.}  When $\Theta \le \thetainf$, the observed \editTwo{specific} flux from the lens is \editThree{$d\Lstar/dE \times \editTwo{\Rlens^2 / (4 \xStar^2)} \times 1/(\pi \thetainf^2 d^2)$}:
\begin{equation}
\frac{dF_{\rm lens}}{dE} = \eta \frac{d\Lstar/dE}{4 \pi d^2} \left(\frac{\Rlens/\Rstar}{1 + \Rlens/\Rstar \times \editTwo{|\bar{E}^2 - E^2|/E}}\right)^2 ,
\end{equation}
where $d$ is the distance between the observer and the source.  If $\Theta \ge \thetalens$, we \editThree{also observe} the flux from the unobscured emitting region itself, $\Fstar = (d\Lstar/dE)/(4 \pi d^2)$.  \editThree{Otherwise ($\Theta < \thetalens$), the emitter is covered by the element, and only the lens contributes to the total observed flux. The excess flux $\Delta d\Fobs/dE$ is found by summing the observed flux from the lens and the unobscured flux from the emitter, and then subtracting the unmodified flux from the emitter $d\Lstar/dE / (4 \pi d^2)$.}  An occultation event produces a change in the observed flux:
\begin{multline}
 \Delta \frac{d\Fobs}{dE}  = \frac{d\Lstar/dE}{4 \pi d^2} \\
                         \times \begin{cases}
                         \displaystyle \editTwo{\eta} \left(\frac{\Rlens/\Rstar}{1 + \Rlens/\Rstar \times |\bar{E}^2 - E^2|/E^2}\right)^2 - 1  & \\
												  & \hspace{-0.75cm}\hspace{-0.5cm}(\Theta \le \thetalens, \thetainf)\\
												 \displaystyle \editTwo{\eta} \left(\frac{\Rlens/\Rstar}{1 + \Rlens/\Rstar \times |\bar{E}^2 - E^2|/E^2}\right)^2 & \\
												  & \hspace{-1.07cm}\hspace{-0.5cm}(\thetalens \le \Theta \le \thetainf)\\
												 -1 & \hspace{-1.07cm}\hspace{-0.5cm}(\thetainf < \Theta \le \thetalens)\\
												 0  & \hspace{-0.75cm}\hspace{-0.5cm}(\thetainf, \thetalens < \Theta)
												 \end{cases}
\end{multline}
At energies near $\bar{E}$, a \editThree{chromatic} lens flare proceeds much like an achromatic lens flare: a dip in the flux as the plate blocks the source, followed by a brilliant flash as the \editThree{element} lenses the source, and a return to the dip as the lens continues to move past the source, before ultimately returning to normal (see the animation in Figure~\ref{fig:SpectrumMovie}).  The intensity of the central flash is lessened and its duration lengthened away from perfect collimation.  At energies far from $\bar{E}$, the \editTwo{specific flux} actually remains in a dip if the lens is directly in front of the source, as the lens weakly glows with intensity $< I_{\star}$ both during the occultation and at surrounding times.

The fluence from the event consists of two contributions, a deficit as the shadow of the lens passes over the observer and blocks the source, and a gain when the observer is in the cone of light that is created by the lens.  \editTwo{The lens blocks the source for a time $\tshadow = \sqrt{\Rlens^2 - b^2} / \vrel$ and transmits light for $\tflare = \sqrt{\thetainf^2 \xStar^2 - b^2} / \vrel$, where $b \equiv \xStar \min \Theta$ is the projected impact parameter of the source.}  Consider the case when the \editThree{source} passes directly \editThree{behind the lens} (\editTwo{$b = 0$}) with constant \editThree{relative} speed \editThree{$\vrel$}.  The lens then blocks the source for a time $\tshadow = \editTwo{2} \Rlens / \editThree{\vrel}$, neglecting the small size of the source, causing a fluence deficit of $d\Fluenceshadow/dE = dF/dE \times \tshadow$:
\begin{equation}
\frac{d\Fluenceshadow}{dE} = \frac{d\Lstar/dE}{4 \pi d^2} \frac{2 \Rlens}{\vrel}.
\end{equation}
We observe transmitted light through the lens for a time $\tflare = 2 \thetainf \xStar / \vrel$, and the fluence of this transmitted light is:
\begin{multline}
\frac{d\Fluenceflare}{dE} = \eta \frac{d\Lstar/dE}{4 \pi d^2} \left(\frac{\Rlens}{\Rstar}\right)^2 \frac{2 \Rstar}{\vrel} \times \\
                            \left(1 + \frac{\Rlens}{\Rstar} \frac{|\bar{E}^2 - E^2|}{E^2}\right)^{-1} .
\end{multline}
So\editTwo{, if the lens center passes exactly over the source,} the net fluence excess caused by this lens flare event is
\begin{multline}
\frac{d\Fluenceexcess}{dE} = \frac{d\Lstar/dE}{4 \pi d^2} \frac{2 \Rlens}{\vrel} \times \\
                             \left(\frac{\eta \Rlens / \Rstar}{1 + (\Rlens/\Rstar) |\bar{E}^2 - E^2|/E^2} - 1\right) .
\end{multline}

Using a \editThree{single lens with chromatic aberration} instead of an achromatic \editThree{system} may be easier and produces a unique spectral evolution that might be recognized by observers that are either nearby or have powerful instruments.  The lens flare event is much dimmer, however, limiting its use as a beacon.  Figure~\ref{fig:SpectrumMovie} shows an example of how the spectrum of an emission region changes during a lensing event.  When the lens is perfectly aligned with the source, the resultant spectrum is strongly peaked (\editThree{thin black} line).  Although it reaches the achromatic specific luminosity at $E = \bar{E}$, the bolometric luminosity of the lens is far less than in the achromatic case (\editThree{solid} grey line), and at low energies, it is less than the unobscured source (\editThree{dashed grey} line).  \editTwo{For the chromatic lens shown in Figure~\ref{fig:SpectrumMovie}, the peak amplification of the bolometric luminosity is $70$ instead of $10^4$.}  The shadow of the lens manifests at $\Theta \ga \thetastar$ as the suppression of \editThree{the flux in} an energy range containing this peak .  The weak emission of the zone plate at low energies, which is visible even when the \editThree{element} does not cover the emission region \editThree{is minor}.

\subsection{Zone plates and other diffractive elements}
The Fresnel zone plate is a very simple optical system capable of forming X-ray images at very long focal lengths.  It consists simply of alternating, concentric opaque and transparent rings, where the ring width decreases further from the center.  Variants like Phase Fresnel Lenses delay the phases of incoming electromagnetic waves instead of absorbing them, and attain a high efficiency \citep{Skinner02,Skinner10}.  The whole disk then has a focal length that depends on the wavelength of the incident light.  

\editThree{Diffractive optics also suffers from chromatic aberration.}  \editThree{The} focal length is $f(E) = \bar{f} E / \bar{E}$, where $\bar{f}$ is some reference focal length for reference energy $\bar{E}$ \citep[e.g.,][]{Skinner02}.  \editTwo{Thus, photons of a particular energy are emitted into a cone with angular radius}
\begin{equation}
\editTwo{\thetainf \approx \frac{\Rstar}{\xStar} \left(1 + \frac{\Rlens}{\Rstar} \frac{|\bar{E} - E|}{E}\right),}
\end{equation}
\editTwo{and the flux and fluence can be calculated similarly as a lens with chromatic aberration.} The observed flux from the \editThree{element itself} is:
\begin{equation}
\frac{dF_{\rm lens}}{dE} = \eta \frac{d\Lstar/dE}{4 \pi d^2} \left(\frac{\Rlens/\Rstar}{1 + \Rlens/\Rstar \times |\bar{E} - E|/E}\right)^2 ,
\end{equation}
\editThree{resulting in a flux change of:}
\begin{multline}
 \Delta \frac{dF_{\rm obs}}{dE}  = \frac{d\Lstar/dE}{4 \pi d^2} \\
                         \times \begin{cases}
                         \displaystyle \editTwo{\eta} \left(\frac{\Rlens/\Rstar}{1 + \Rlens/\Rstar \times |\bar{E} - E|/E}\right)^2 - 1  & \\
												  & \hspace{-0.75cm}(\Theta \le \thetalens, \thetainf)\\
												 \displaystyle \editTwo{\eta} \left(\frac{\Rlens/\Rstar}{1 + \Rlens/\Rstar \times |\bar{E} - E|/E}\right)^2 & \\
												  & \hspace{-1.07cm}(\thetalens \le \Theta \le \thetainf)\\
												 -1 & \hspace{-1.07cm}(\thetainf < \Theta \le \thetalens)\\
												 0  & \hspace{-0.75cm}(\thetainf, \thetalens < \Theta)
												 \end{cases}
\end{multline}
\editThree{The net fluence of the event is:}
\begin{multline}
\frac{d\Fluenceexcess}{dE} = \frac{d\Lstar/dE}{4 \pi d^2} \frac{2 \Rlens}{\vrel} \times \\
                             \left(\frac{\eta \Rlens / \Rstar}{1 + (\Rlens/\Rstar) |\bar{E} - E|/E} - 1\right) \editTwo{,}
\end{multline}
\editTwo{if the source is small and it passes directly behind the lens ($b = 0$).}

\editThree{The time-dependent spectrum observed when a diffractive element passes directly in front of the NS is illustrated in Figure~\ref{fig:SpectrumMovie}.  Qualitatively, it behaves similarly to a single refractive lens with chromatic aberration, but its spectral peak is wider and the event is brighter since $f$ has a weaker dependence on $E$.} \editTwo{It attains a maximum bolometric amplification of $130$.}

\subsection{Lens flare ``beacons'' for isolated neutron stars}
\editThree{An isolated NS may be considered as stationary relative to the orbits of lenses.}  If ETIs want to set up an isotropic beacon to draw the attention of others in all directions, they need many lenses in orbit of the LMXB to ensure that at least one passes in front of the NS as viewed from each direction.  Furthermore, a recipient with relatively poor X-ray detection abilities, like ourselves, is unlikely to detect individual eclipses at large distances because eclipses are so short and the photon fluence deficit is so small.\footnote{Otherwise, ETIs could settle for using a simple occulter rather than an X-ray lens, or even use a partial Dyson sphere/swarm as envisioned in \citealt{Chennamangalam15} and \citealt{Imara18}.}  Thus, they would have to ensure that at least one lens transits with impact parameter $<\Rstar$ of the NS as viewed from each direction \editTwo{to ensure flares of peak luminosity are observed}.  This requires a swarm of at least $\Nlens \ga \pi \xStar / \Rstar$ lenses. \editThree{A lens flare then occurs once per orbital period.}

\editThree{The ETIs may want to ensure a higher lens flare rate if the lenses have very long orbital periods $\Plens$, with correspondingly large $\alens$ (assumed equal to $\xStar$).  To ensure a flare rate of $\Gammaflare$ ($>1/\Plens$) \editTwo{as defined by passages within $\Rstar$}, the necessary number of lenses is $\pi \alens/\Rstar \times \Plens \Gammaflare$:}
\begin{equation}
\Nlens \approx \frac{2 \pi^2 \Gammaflare}{\Rstar} \sqrt{\frac{\alens^5}{(1 - \beta) G \MNS}} ,
\end{equation}
\editThree{where $\MNS$ is the NS mass, $G$ is Newton's gravitational constant, and $1 - \beta$ is a factor that parameterizes the effect of radiation pressure (Section~\ref{sec:Luminosity}).  For $\alens \sim 30\ \AU$ and $\Gammaflare \sim 1/(30\ \yr)$, $\sim 10^{10}$ lenses are necessary.}

\editThree{The flare duration $\tflare$ is determined by the lens orbital speed, and is $\Rstar / \vlens \approx \Rstar \sqrt{\alens/[(1 - \beta) G \MNS]}$:}
\begin{multline}
\tflare \approx 1.6\ \sec\, \left(\frac{\Rstar}{10\ \km}\right) \left(\frac{\alens}{30\ \AU}\right)^{1/2} \\
\times \left(\frac{(1 - \beta) \MNS}{1.4\ \Msun}\right)^{-1/2} .
\end{multline}

\subsection{Lens flare ``beacons'' in binary systems}
\label{sec:Binary}
\editThree{LMXBs are binary systems, and thus the neutron star itself moves.  In a typical LMXB system, the donor is a low mass dwarf or possibly subgiant with mass $\Mdonor \sim 0.5\ \Msun$, with an orbital period $\Pstar$ of several hours to one day \citep[e.g.,][]{Pfahl03}.  From most inclinations, the NS traces out an elliptical path on the sky as viewed by the observer, with a projected semimajor axis $\editTwo{\aNS}$ of $\la 1\ \Rsun$.  The NS motion therefore is more important than the lens's orbital motion.}

\editThree{In the binary case, a large number of lens orbits pass over the NS orbit from the perspective of the observer.  Typically, the NS will be at the wrong orbital phase to be occulted when a lens crosses its orbit.  The probability of a lens flare during any one lens passage is set by the ratio of the time window for the flare and the NS orbital period:}
\begin{equation}
\Pflare \approx \frac{2 \Rstar}{\vlens \Pbin} \approx \frac{\Rstar}{\pi} \sqrt{\frac{\alens}{(1 - \beta) \abin^3}},
\end{equation}
\editThree{which is of order $10^{-5} \endash 10^{-4}$\editTwo{, where $\abin = \aNS \Mbin/\Mdonor$.}  The probability of a shadow event is $\Rlens/\Rstar$ times bigger.  On the other hand, only one lens needs to cross the entire $\sim \editTwo{\aNS}$ width of the orbit to ensure that a lens flare eventually occurs, unlike the stationary source case where one lens per $\sim \Rstar$ to ensure a lens flare.}

\editThree{Many fewer lenses are needed to achieve an observed \editTwo{peak luminosity} flare rate $\Gammaflare$ when the source is in a binary:}
\begin{equation}
\label{eqn:Nlens}
\Nlens \approx \frac{\pi \alens}{\aNS} \frac{\Plens \Gammaflare}{\Pflare} \approx \frac{2 \pi^3 \alens^2 \Gammaflare}{\aNS \Rstar} \sqrt{\frac{\abin^3}{G \Mbin}}.
\end{equation}
\editThree{For $\alens \sim 30\ \AU$, $\editTwo{\aNS} \sim \Rsun$, and $\Gammaflare \sim 1/(30\ \yr)$, about $2 \times 10^9$ lenses are needed.  Flare events are shorter because the NS rapidly crosses the center of the lens in a time $\Rstar (\Mbin/\Mdonor) \sqrt{\abin/(G\Mbin)}$:}
\begin{multline}
\tflare \approx 0.13\ \sec\, \left(\frac{\Rstar}{10\ \km}\right) \left(\frac{\aNS}{\Rsun}\right)^{1/2} \left(\frac{\Mbin}{1.9\ \Msun}\right) \\
\times \left(\frac{\Mdonor}{0.5\ \Msun}\right)^{-3/2} .
\end{multline}

\editThree{A further consideration is that the NS orbit requires a minimum field-of-view $\aNS / \alens$, set by the distance between the NS and the barycenter, assuming the lens has synchronous rotation.  Diffraction-based lenses like zone plates have narrow fields of view, especially severe for huge zone plates in hard X-rays.  Field curvature and astigmatism imply aberration-free fields of view of just $\sqrt{f \lambda}/\Rlens$, and coma limits the aberration-free field to $\lambda f^2 / (2 \Rlens^3)$ \citep{Young72}.  A zone plate must therefore have a minimum orbital distance of}
\begin{multline}
\alens > \left(\frac{\editTwo{\aNS}^2 \Rlens^2 E}{hc}\right)^{1/3} > 105\ \AU\ \left(\frac{\aNS}{\Rsun}\right)^{2/3} \\
\times \left(\frac{\Rlens}{1,000\ \km}\right)^{2/3}  \left(\frac{E}{10\ \keV}\right)^{1/3} .
\end{multline}
\editThree{Because of the extremely low tolerance for misaim, it may be easier to replace a singular plate with many smaller elements.  These would require wedge prisms to align light before entering the smaller zone plates, which themselves would introduce severe chromatic aberration.}

\subsection{Source luminosity constraints}
\label{sec:Luminosity}
The extreme luminosity of the LMXB necessitates that the lenses be placed very far from the binary to avoid overheating and sublimating.  This is especially true for zone plates, which absorb a large fraction of the luminosity incident on them.  If a lens is maintained at temperature $\Tlens$ around a\editTwo{n} LMXB of luminosity $\Lstar$, and it absorbs a fraction $\zetaabs$ of incident flux, its distance from the LMXB must be $\xStar = \sqrt{(\zetaabs \Lstar)/(4 \pi \sigma_{\rm SB} \Tlens^4)}$:
\begin{multline}
\label{eqn:LensOrbitVsTemp}
\xStar = 6.3\ \AU\, \left(\frac{\zetaabs \Lstar}{10^{38}\ \erg\,\sec^{-1}}\right)^{1/2} \left(\frac{\Tlens}{2,000\ \Kelv}\right)^{-2}.
\end{multline}
\editThree{In principle, $\zetaabs$ might be very small, as might be achieved with phase Fresnel lenses or phase zone plates \citep{Skinner10}.}

\editThree{Practical considerations may require the lenses to be placed much further still.  Because LMXB luminosities are variable on multiple timescales \citep[e.g.,][]{vanDerKlis89}, the lens will have an erratic temperature, heating and cooling as the LMXB fluctuates.  The structures therefore will expand and contract.  In order to reduce the enormous stresses, they will then need expansion joints or some other solution.  The thermal fluctuation in the lens size might also change the focal length of the lens.} \edit1{Thermal stresses also pose a threat to the precise engineering needed for an optical element, causing distortions in the lens shape or misalignment of sublenses.}

\editThree{Radiation pressure from a bright LMXB is also a challenge.  The ratio of the radiation pressure force and gravitational force on the lens is:}
\begin{multline}
\beta = \frac{\zetaabs \Lstar}{4 \pi G \Mbin \Sigmalens c} = 0.10 \left(\frac{\zetaabs \Lstar}{10^{38}\ \erg\,\sec^{-1}}\right) \\
\times \left(\frac{\Mbin}{1.9\ \Msun}\right)^{-1} \left(\frac{\Sigmalens}{10\ \gcm2}\right)^{-1} .
\end{multline}
\editThree{where $\Sigmalens$ is the projected surface density of the entire lens structure.  The X-ray optical depth of a material is generally set by its composition and surface density.  For optically thin materials, $\zetaabs \propto \Sigmalens$, and the energy loss grammage of 10 keV photons is} $\sim 3\ \gcm2$ for lithium, $0.4\ \gcm2$ for carbon, $0.03\ \gcm2$ for silicon, and $0.005 \endash 0.01\ \gcm2$ for iron and heavier metals \citep{Hubbell96,Tanabashi18}. \editThree{Note that a significant fraction of the LMXB luminosity is the softer emission from the accretion disk, which should couple even more efficiently to the structure.  Large thin plates made of these materials would be blown away by the radiation pressure.  The problem is particularly acute for ULXs.}

\editThree{If the X-ray source is steady, $\beta \le 1$ might be sufficient: the lenses would then be levitated by radiation pressure and could move very slowly in their orbits as ``quasites'' \citep{Kipping19}.  The fluctuating radiation pressure from bright LMXBs would cause erratic changes to the orbits of the lenses, however.  Therefore, practical lens structures around the brightest LMXBs need to be weighted down.  The ``ballast'' could consist of long, thin pegs with minimal cross section scattered in the lens.  Additional mass is likely to be contributed by supporting structures for the lens, as well as counterweights needed to ensure rotational stability (see Appendix~\ref{sec:RotationalStability}).  The problem may also be somewhat ameliorated if thermal emission from the lens can be confined to a small cone in the direction opposite the LMXB, to push the lens back, although the practicality depends on the lens temperature.}

\subsection{Mass requirements}
\label{sec:Mass}
The lenses can be thought of as thin structures, with a thickness of only a few centimeters, despite being hundreds of kilometers wide.  \editThree{To this is added the mass of support structures, counterweights, and ballast.}  \editThree{The} total mass of a single lens is $\pi \Rlens^2 \Sigmalens = \pi \AreaFactor \Rstar^2 \Sigmalens$:
\begin{multline}
\Mlens \approx 3.1 \times 10^{17}\ \gram\ \left(\frac{\Rlens}{1,000\ \km}\right)^2 \left(\frac{\Rstar}{10\ \km}\right)^2 \\
\times \left(\frac{\Sigmalens}{10\ \gram\,\cm^{-2}}\right).
\end{multline}
Individually, this is quite small compared to other proposed megastructures -- \editThree{hundreds} of gigatons per lens, about the mass of an asteroid with a radius of \editThree{three kilometers}.  It is especially impressive when one considers that this artifact is capable of repeatedly producing a signal \editThree{potentially} observable at intergalactic distances.

In order for lens flares of peak brightness to be visible from any direction, \editThree{millions of lenses are needed, and billions are needed to ensure they occur frequently (section~\ref{sec:Binary}).}  The total mass of all the lenses is \editThree{$\Mswarm = \Nlens \Mlens$}:
\begin{multline}
\Mswarm \approx \frac{2 \pi^4 \alens^2 \Rlens^2 \Sigmalens \Gammaflare}{\aNS \Rstar} \sqrt{\frac{\abin^3}{G \Mbin}} \\
        \approx 5.1 \times 10^{26}\ \gram\ \left(\frac{\alens}{30\ \AU}\right)^2 \left(\frac{\Rlens}{1,000\ \km}\right)^2 \\
				\times \left(\frac{\Sigmalens}{10\ \gram\,\cm^{-2}}\right)  \left(\frac{\Gammaflare}{(30\ \yr)^{-1}}\right) \left(\frac{\aNS}{\Rsun}\right) \left(\frac{\Rstar}{10\ \km}\right)^{-1} \\
				\times \left(\frac{\Mbin}{1.9\ \Msun}\right)  \left(\frac{\Mdonor}{0.5\ \Msun}\right)^{-3/2} .
\end{multline}
This is about the mass of \editThree{Mars}.  Simple occulting systems have much less onerous requirements, since even an off-center eclipse blocks the central source effectively, reducing the necessary mass by $\Rstar/\Roccult$. 

 \editThree{The} material requirements are likely to be among the greatest challenges of building this system.  Any solid materials originally present in the system face a double onslaught: first, from the supernova that accompanied the birth of the neutron star, and second, from the LMXB luminosity itself.  Even if solid planets survived a supernova or formed from the remnant, an Eddington luminosity LMXB would sublimate all rocky material within about ten AU \citep{Miller01}.  Asteroid belts without planets around pulsars, including millisecond pulsars, have been proposed to explain timing anomalies, nulling, and radio transients \citep[as in][]{Cordes08,Campana11,Mottez13,Shannon13,Brook14,Huang14,Geng15}.  These could be mined if they are present during the LMXB phase.  

The planetary system around the millisecond pulsar PSR B1257+12 is possible evidence that X-ray binaries can host solid materials \citep{Wolszczan92}, as the pulsar may have passed through an X-ray binary phase during its spin-up \citep{Alpar82}.  Many theories as to how these mysterious planets formed have been advanced \citep{Phinney93,Podsiadlowski93,Martin16}.  While it remains possible the planets formed out of the supernova remnant through a fallback disk \citep[e.g.,][]{Lin91,Currie07,Hansen09}, the planets may instead have formed from the disruption of a companion star, which would only happen after (or at the end of) any LMXB phase \citep[e.g.,][]{Stevens92,Rasio92,Banit93,Martin16,Margalit17}.  \citet{Tavani92} suggested PSR B1257+12's planets formed during an LMXB phase, with the accretion disk acting as a shield against the intense X-ray emission.  \citet{Miller01} instead hypothesized that PSR B1257+12 was born a millisecond pulsar, and so avoided any binary phase.   If LMXB planets do exist, they might be found by observing their transits \citep{Imara18}.  

The prospective engineers would be left with few \editThree{other} choices.  They might import material from distances $\ga 10\ \AU$ -- from surviving distant planets or Oort clouds, post-supernova Kuiper belts formed from fallback disks, or even from other star systems.  \editThree{The Solar Oort Cloud's mass has been estimated at \editTwo{one to} tens of Earth masses, for example, most within an inner cloud a few thousand AUs out \citep{Weissman96,Boe19}.}  If a fallback disk with rocky material forms, they may also build the lenses after the supernova but before the LMXB achieves full luminosity, though it could be millions of years before it became useful.  Planets around high mass stars remain unconstrained, and fallback disks and pulsar planets seem to be rare \citep{Wang14,Kerr15}.  On the other hand, transporting \editThree{$\sim 10^{27}\ \gram$} across interstellar distances requires vast amounts of energy and/or time.  A more radical option would be to mine the accretion disk or donor star itself.  They might more easily mine planetesimals in a circumbinary disk during the LMXB phase, if they exist \citep{Tavani92}.

Another possibility is to use a much less luminous X-ray emitting system as the light source\editThree{, reducing temperature and radiation pressure constraints (Section~\ref{sec:Luminosity})}.  Fainter LMXBs number in the hundreds in the Milky Way \citep{Grimm02}.  High-mass X-ray binaries with neutron stars in the Milky Way also have luminosities less than $3 \times 10^{37}\ \erg\,\sec^{-1}$ (with fainter ones being more abundant), but they are rarer than LMXBs \citep{Grimm02}.  Non-accreting NSs emit thermal X-rays, as they maintain temperatures of $\sim 10^6\ \Kelv$ for about 1 Myr \citep[e.g.,][]{Pavlov02,Yakovlev04}.  Older NSs of ages $>1\ \Myr$ may have rotationally-powered X-rays, apparently from small hotspots, though the luminosities are low, around $10^{28} \endash 10^{34}\ \erg\,\sec^{-1}$ \citep[as in][]{Pavlov09,Posselt12}.  They are far less practical as beacons visible across the Galaxy or between galaxies.

\subsection{Additional challenges}
Although the lenses do not require power to continue beaming the LMXB luminosity, the swarm presents additional challenges that could require maintenance.  

First, it is necessary to ensure that the lenses do not crash into one another.  Suppose the lenses are distributed randomly in a shell with thickness \editThree{$\Delta_a \alens \ll \alens$} and have random velocities \editThree{$\vrand$}.  With a collisional cross section area \editThree{$\zetageom \pi \Rlens^2$}, the rate of collisions is \editThree{$\Gammacoll \approx \Nlens \zetageom \Rlens^2 / (4 \Delta_a \alens^3) \times \sqrt{(1 - \beta) G \Mbin / \alens} \times (\vrand / \vlens)$}.  \editThree{From equation~\ref{eqn:LensOrbitVsTemp} relating lens orbital distance to temperature}{From equation~\ref{eqn:Nlens}}:
\begin{multline}
\Gammacoll \approx \frac{\pi^3 \zetageom \Rlens^2 \Gammaflare}{2 \Delta_a \aNS \Rstar} \sqrt{(1 - \beta) \frac{\abin^3}{\alens^3}} \left(\frac{\vrand}{\vlens}\right) \\
           \approx (\editTwo{94\ \kyr})^{-1} \zetageom \sqrt{1 - \beta} \left(\frac{\Rlens}{1,000\ \km}\right)^2 \left(\frac{\Gammaflare}{(30\ \yr)^{-1}}\right) \\
					\times \left(\frac{\Delta_a}{0.1}\right)^{-1} \left(\frac{\aNS}{\Rsun}\right)^{1/2} \left(\frac{\Rstar}{10\ \km}\right)^{-1} \left(\frac{\alens}{30\ \AU}\right)^{-3/2} \\ 
					\times \left(\frac{\Mbin}{1.9\ \Msun}\right)^{3/2} \left(\frac{\Mdonor}{0.5\ \Msun}\right)^{-3/2} \left(\frac{\vrand}{\vlens}\right)^{-1} .
\end{multline}
\editThree{Actually, because each collision produces many pieces of potentially hazardous debris, the timescale for a collisional cascade could be one or two orders of magnitude shorter (\citealt{Kessler78}; see Appendix~\ref{sec:CollisionalCascades}).}

\editThree{Destruction by a collisional cascade may be inhibited by not placing the lenses on random orbits.}  Lenses might be placed at different distances from the LMXB, each with an appropriate focal length, to spread out the space between them.  Furthermore, subsets of lenses with different orbital planes can have highly ordered velocities, with minimal dispersions, so encounter speeds are much slower.  \editThree{The small dispersions both reduce the rate and severity of collisions.}  The true lifetime of a lens swarm may be determined by how quickly the lens orbits are perturbed, or are subject to its own gravitational instabilities.  \editThree{I estimate the relevant timescales due to the quadrupole field of the binary, Jeans instability, and lens-lens encounters in Appendix~\ref{sec:OrbitalStability}, finding they take many millennia to play out.  Fluctuating radiation pressure is likely to randomize the orbits at least somewhat, although radiation pressure also blows out debris from collisions.}  This traffic control problem is a frequent one that arises in the consideration of ETI megastructures \citep{Carrigan09,Lacki16-K3,Sallmen19}.  \editThree{Alternatively, the collisional lifespan may be extended by placing lenses even further from the NS at the cost of more lenses, or by using smaller lenses.}  Simple occulters also have a reduced collision problem because fewer are needed.  

Guidance may also be necessary to ensure the lenses collimate the LMXB optimally.  \editThree{At the very least, a counterweight in the form of a central pole or enclosing tube is needed to ensure the rotational stability of the lens (see Appendix~\ref{sec:RotationalStability}).  If the orbit of the lens attains an eccentricity of $e$, then even a synchronously rotating lens will not be pointed direct at the system barycenter, but will deviate by up to $2e$ radians.  Thus, the lenses either need to maintain very low eccentricity, or have relatively wide fields-of-view.}  \editThree{The limited depth of field is another reason for large, achromatic lenses to maintain circular orbits.}  This issue is not a problem for \editThree{elements that} already suffer from chromatic aberration, with different photon energies already having different focal lengths. \editThree{Collimators with many small lenses have much greater tolerance for radial drift.}

\edit1{Finally, the fabrication of the lenses would require precise shaping on scales far beyond our capabilities.  Even small deviations in the lens or zone element shape would degrade performance, or even suppress the focusing ability entirely.  For example, a single zone plate with radius $1,000\ \km$ and focal length $30\ \AU$ at $10\ \keV$ would have about two billion zones, the outermost with width $0.03\ \cm$.  If the assembly actually consists of many smaller zone plates, the requirements are less severe: a zone plate subplate with radius $0.1\ \km$ only needs about twenty zones, the outermost with width $300\ \cm$.  However, the output beams from these subplates need to be aligned within an angle $\la \thetastar \approx 0.5\ \mas (\Rstar / 10\ \km) (f / 30\ \AU)^{-1}$, equivalent to a deviation of less than $\sim 2\ \um\ \km^{-1} (\Rstar / 10\ \km) (f / 30\ \AU)^{-1}$.  This may require the subelements to be shifted slightly over time using some internal system.  The extreme thermal stresses likely induced by variability of bright LMXBs makes this an especially difficult challenge.  Of course, the structure should still act as an occulter even if the optical system fails, but the resultant signal would be harder to detect and more ambiguous.}

\section{Observing Lens Flares}
\label{sec:ObservingLensFlares}
\subsection{Prospects with X-ray facilities}
How effective are LMXB lenses as beacons?  Could our instruments detect them across \editThree{the Galaxy, or even at} intergalactic distances?

During a detectable lens flare, a number of photons arrive nearly simultaneously from the LMXB.  The near-coincidence of their arrival times would stand out compared to the relatively steady background flux.  In addition, there's a much smaller photon deficit during the shadow phase of the occultation.  Both the number of photons and their coincidence are diluted if the \editThree{optical element has chromatic aberration}.  \editThree{It's extremely unlikely, however, that a lens flare would occur while a narrow field detector happened to observe a given LMXB.  For this reason, I \editTwo{shall consider observability with} several wide-field detectors that observe $\ga 1\ \sr$ at a time \editTwo{in addition to the more sensitive narrow-field instruments}.  These instruments, generally designed to locate gamma-ray bursts, have effective areas that can be $\sim 10 \endash 100$ times smaller than sensitive narrow-field instruments.}

\editTwo{The signal-to-noise ratio is calculated for a lens flare event with a source passing directly behind the lens ($b = 0$).  I also make the simplifying assumption that the source is small compared to the lens, with all parts of the source being occluded or magnified simultaneously.}  For a detector with effective area $\Aeff (E)$, \editThree{the photon excess detected during a time window $\twindow$} is:
\begin{multline}
\Delta \NBL = \editTwo{\int \left[\frac{d\Fluxflare}{dE} \min\left(\twindow, \frac{2 \thetainf(E) \xStar}{\vrel}\right) \right.}\\
\editTwo{\left. - \frac{d\Fluxshadow}{dE} \min\left(\twindow, \frac{2 \Rlens}{\vrel}\right) \right] \frac{\Aeff (E)}{E} dE }, 
\end{multline}
where the excess fluence spectrum follows from the luminosity spectrum and the lens characteristics.  \editThree{I consider two window lengths, \editTwo{one for the ``peak'' of the flare and one for the occultation length.  After taking into account the minimum time resolution $\tres$ of the instrument, these are $\twindowflare = \min(\tflare, \tres)$ and $\twindowshadow = \min(\tshadow, \tres)$.}  There is only a miniscule difference in the case of an achromatic lens, but the peak accounts for only a fraction of the excess fluence for zone plates.}    

\editTwo{The boundary layer of a highly accreting LMXB} is Compton-opaque and has a Wien spectrum \citep{Popham01,Gilfanov03}.  From \citet{Rybicki79}:
\begin{equation}
\frac{d\LBL}{dE} = \frac{\LBL E^3}{6 (k_B \Tstar)^4} \exp\left(-\frac{E}{k_B T_{\rm BL}}\right) .
\end{equation}  
The temperature of the boundary layer has a common mean temperature of $k_B T_{\rm BL} = 2.4\ \keV$ \citep{Gilfanov03,Revnivtsev13}. \editThree{The unlensed flux from the accretion disk serves as \editTwo{an unavoidable} background.  I use the multicolor model to estimate its observed spectrum:}
\begin{equation}
\frac{d\LDisk}{dE} \propto \int^{\Tdiskin}_{0} \frac{E^3}{\displaystyle \exp[E/(k_B T)] - 1} \left(\frac{T}{\Tdiskin}\right)^{-11/3} \frac{dT}{\Tdiskin}, 
\end{equation}
\editThree{where $k \Tdiskin = 1.4\ \keV$ is typical of LMXBs and the spectrum is normalized to an apparent luminosity $\LDisk$ \citep{Mitsuda84}.}

\editTwo{Wide-field X-ray instruments without focusing optics, have high background rates because background events from many parts of the sky are confused with those from the source.  X-ray instruments with focusing optics generally have low background rates because only those falling within the point-spread-function of the source contribute to the noise.  The background rate $\dot{\Nback}$ for each considered instruments is listed in Table~\ref{table:SNR}; it dominates over the normal flux from the BL by one to three orders of magnitude.}

\editThree{I then estimate the expected signal-to-noise ratio of the event from $\Delta \NBL$ detected and the effective background count $\Nback \editTwo{= \dot{\Nback} \twindow}$:
\begin{equation}
\SNR = \frac{|\Delta \NBL|}{\sqrt{\NBLnorm + \max(\Delta \NBL, 0) + \Ndisk + \Nback}},
\end{equation}
where $\NBLnorm$ is the expected number of photons detected from the boundary layer when no event occurs and $\Ndisk$ is the expected number from the accretion disk \editTwo{during the window}.}

\begin{deluxetable*}{llcccccccccc}
\tabletypesize{\scriptsize}
\tablewidth{0pt}
\tablecolumns{12}
\tablecaption{Signal-to-noise during lens flare \label{table:SNR}}
\tablehead{\colhead{Facility} & \colhead{Instrument} & \colhead{$\tres$} & \colhead{$\dot{\NBLnorm} (\sec^{-1})$} & \colhead{$\dot{\Ndisk} (\sec^{-1})$} & \colhead{$\dot{\Nback} (\sec^{-1})$} & \colhead{Shadow} & \colhead{Achromatic} & \multicolumn{2}{c}{Chromatic lens} & \multicolumn{2}{c}{Diffractive element} \\ & & & & & & \colhead{$\twindowshadow$} & \colhead{$\twindowflare$} & \colhead{$\twindowflare$} & \colhead{$\twindowshadow$} & \colhead{$\twindowflare$} & \colhead{$\twindowshadow$}}
\startdata
\cutinhead{Wide-field instruments}
HETE-2                         & FREGATE\tablenotemark{\ensuremath{\star}}  & $6.4\ \usec$  & $66$    & $26$     & $2,200$   & $6.9$   & $410$   & $32$    & $22$     & $49$    & $39$\\
                               & WXM\tablenotemark{\ensuremath{\star}}      & $256\ \usec$  & $50$    & $49$     & $530$     & $10$    & $350$   & $30$    & $26$     & $43$    & $46$\\RHESSI\tablenotemark{\ensuremath{\star}} &                                  & $100\ \msec$  & $11$    & $1.8$    & $1,800$   & $1.3$   & $160$   & $8.0$   & $3.8$    & $14$    & $7.0$\\
\emph{Swift}                   & BAT                                        & $140\ \usec$  & $40$    & $0.4$    & $40,550$  & $1.0$   & $300$   & $0.05$  & $0.2$    & $0.2$   & $0.7$\\
\cutinhead{Intermediate-field instruments}
MAXI                           & GSC                                        & $100\ \usec$  & $5.9$   & $6.6$    & $120$     & $2.6$   & $120$   & $9.5$   & $6.6$    & $14$    & $12$\\
\cutinhead{Narrow-field instruments}
\emph{Athena}                  & WFI\tablenotemark{\ensuremath{\star\star}} & $450\ \usec$  & $2,840$ & $36,435$ & $0.001$   & $72$    & $2,700$ & $140$   & $55$     & $220$   & $160$\\
\emph{Chandra}                 & ACIS-I                                     & $3.2\ \sec$   & $133$   & $717$    & 1E-6      & $23$    & $580$   & $52$    & $16$     & $81$    & $47$\\
                               & HRC-I                                      & $16\ \usec$   & $25$    & $375$    & 5E-6      & $6.2$ & $250$ & $13$ & $4.6$  & $20.0$  & $13$\\
\emph{eROSITA}                 &                                            & $50\ \msec$   & $125$   & $2,863$  & $0.00485$ & $11$    & $560$   & $16$    & $1.1$    & $27$    & $9.3$\\
NICER                          &                                            & $100\ \nsec$  & $328$   & $3,736$  & $0.17$    & $26$    & $910$   & $45$    & $15$     & $69$    & $49$\\
\emph{NuSTAR}                  &                                            & $2\ \usec$    & $444$   & $277$    & $0.001$   & $83$    & $1,100$ & $99$    & $150$    & $140$   & $230$\\
RXTE                           & PCA\tablenotemark{\ensuremath{\star}}      & $1\ \usec$    & $3,244$ & $2,401$  & $96$      & $210$   & $2,800$ & $260$   & $390$    & $370$   & $600$\\
\emph{XMM-Newton}              & EPIC pnCCD                                 & $73.3\ \msec$ & $477$   & $3,170$  & $0.0002$  & $39$    & $1,100$ & $87$    & $73$     & $130$   & $140$
\enddata
\tablecomments{Assumed parameters: $\LBL = 10^{38}\ \erg\sec^{-1}$, $k_B \TBL = 2.4\ \keV$, $\Rstar = 10\ \km$, $\LDisk = 10^{38}\ \erg\ \sec{-1}$, $k_B \Tdiskin = 1.4\ \keV$, $\vrel = 80\ \kms$, $\eta = 1$, $\Rlens = 1,000\ \km$, $\bar{E} = 3 k_B \TBL$, $d = 10\ \kpc$, $b = 0$.  Used window listed under element type.}
\tablenotetext{\ensuremath{\star}}{Defunct instrument}
\tablenotetext{\ensuremath{\star\star}}{\hspace{0.15cm}Planned instrument}
\tablerefs{\emph{Athena}: \citet{Barcons15} ($\Aeff$), \citet{Rau13} Figure 10 ($\Nback$, $\tres$); \emph{Chandra} ACIS-I: \citet{Schwartz14} ($\Aeff$), \citet{Garmire03} ($\Nback$, $\tres$); \emph{Chandra} HRC-I: \citet{Schwartz14} ($\Aeff$), \citet{Weisskopf03} ($\Nback$, $\tres$); \emph{eROSITA}: \citet{Merloni12} ($\Nback$ summed particle and photon background over square arcminute); HETE-2 FREGATE: \citet{Atteia03}, $\Nback$ sum of detectors A and B; HETE-2 WXM: \citet{Shirasaki03};  MAXI GSC: \citet{Matsuoka09} ($\Aeff$ from detection efficiency, scaled to $10\ \cm^2$ from \citealt{Sugizaki11}; background rate from idealized $12 \times 10\ \sec^{-1}$), NICER: \citet{Arzoumanian14} ($\Aeff$), \citet{Gendreau12} ($\Nback$ from $0.05\ \sec^{-1}$ plus $12\ \keV \times 0.01\ \sec^{-1}\ \keV^{-1}$ from particles, $\tres$); \emph{NuSTAR}: \citet{Harrison13}; RHESSI: \citet{Lin02} ($\Aeff$, $\tres$), \citet{Smith02} Figure 9 ($\Nback$, integrated to 50 keV); RXTE PCA: \citet{Jahoda96} ($\Nback$ summed over all bands in Table 5), \citet{Bradt93} ($\tres$); \emph{Swift}-BAT: \citet{SwiftBAT05} ($\Aeff$, $\tres$), \citet{Baumgartner13} ($\Nback$, estimated from equation 8); \emph{XMM-Newton}: \citet{XMMNewtonSOC18} ($\Aeff$), \citet{Struder01} ($\Nback$, $\tres$).}
\end{deluxetable*}

I present calculated \editThree{$\SNR$} for several X-ray instruments in Table~\ref{table:SNR}.  Of course, the  actual number detected depends on a variety of parameters, especially $\Rlens$\editTwo{, the distance,} and the properties of the LMXB boundary layer.  I assumed that the lens system is in a bright LMXB system with boundary layer luminosity $\LBL = 10^{38}\ \erg\,\sec^{-1}$ \editThree{and an equal accretion disk apparent luminosity}.  The Galaxy hosts two neutron star LMXBs with total lumnosity greater than $2 \times 10^{38}\ \erg\,\sec^{-1}$ \citep{Grimm02}\footnote{SS433 may in fact be a ULX that beams its emission away from us, but from our point of view it is X-ray faint \citep{Begelman06}.}: Cir X-1 \citep{Linares10} and Sco X-1 \citep[e.g.,][]{MataSanchez15}.\footnote{\citet{Revnivtsev06} find the boundary layer fraction of Sco X-1 varies from $30\% \endash 50\%$.}  When the luminosity is this high, the boundary layer is expected to cover much of the NS \citep{Inogamov99,Popham01}, so I adopt $\Rstar = 10\ \km$.  I also use $\Rlens = 1,000\ \km$, which is generally large enough to ensure a flare from \editThree{a chromatic lens is detectable at the fiducial distance of $10\ \kpc$}, while remaining significantly smaller than the planet-size structures of \citet{Arnold05}.

Efficient ($\eta = 1$) achromatic lenses this big \editThree{around a bright LMXB} produce flares that are \editThree{easily detectable throughout the Galaxy}.  \editThree{After all, the apparent luminosity of the fiducial achromatic lens is comparable to a soft gamma repeater flare, although it lasts only a fraction of a second.  I find signal-to-noise ratios of excess of \editTwo{$120$} for all considered instruments.  \editTwo{They could also be detected by all considered instruments within the Large Magellanic Cloud, at a distance of $49.45\ \kpc$, with $\SNR \ge 24$ \citep{Pietrzynski19}.}  In addition, these flares could be detected by all considered narrow-field instruments with $\SNR \ge 5$ even in M31, $785\ \kpc$ away \citep{McConnachie05}.  \editTwo{ATHENA-WFI and RXTE-PCA could make a $\SNR \sim 7$ detection from M81 ($3.63\ \Mpc$; \citealt{Karachentsev02}) and other galaxies like NGC 5128, NGC 253, and M82.}}

\begin{figure*}
\centerline{\includegraphics[width=8cm]{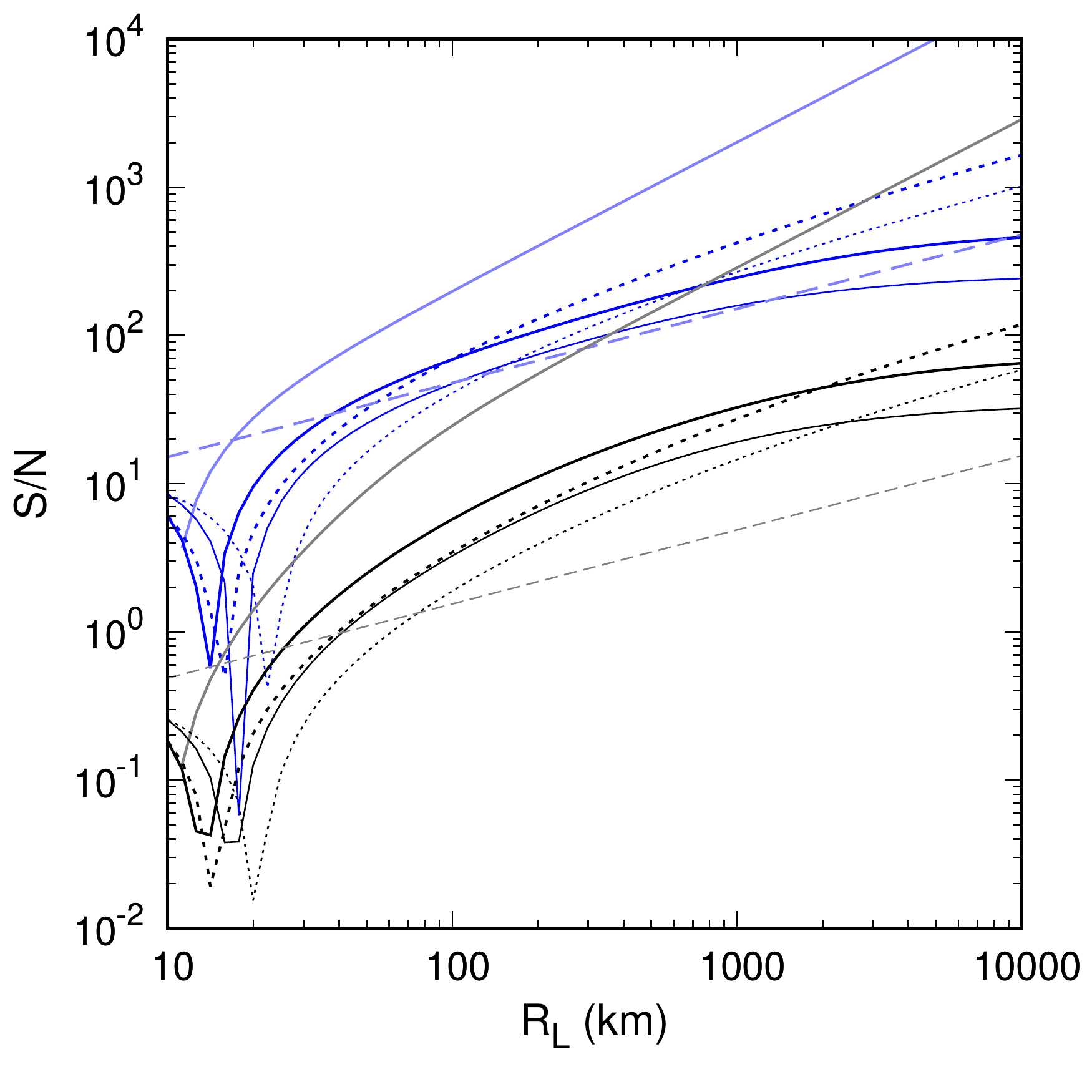}\includegraphics[width=8cm]{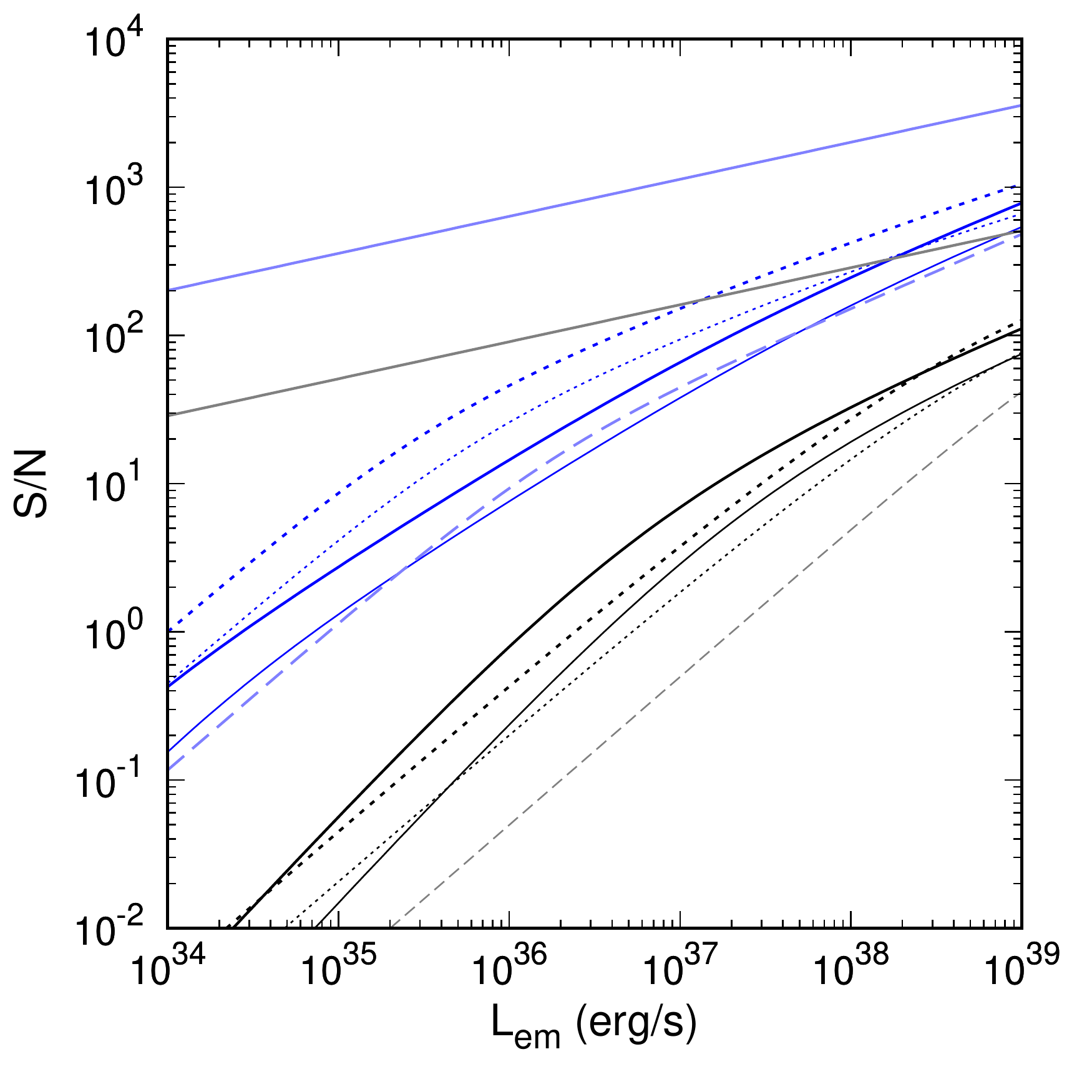}}
\figcaption{\editThree{Signal-to-noise ratio of} a lens flare by RXTE PCA (blue) and HETE-2 FREGATE (black/grey) as a function of lens radius \editTwo{(left) and LMXB luminosity (right; $\Rstar \propto \sqrt{\Lstar}$)}.  The LMXB is assumed to be $10\ \kpc$ away.  \editTwo{The models are calculated under the assumption that $\Rstar \ll \Rlens$, and thus become inaccurate for $\Rlens \sim 10\ \km$.}  Excess counts for an achromatic lens use the solid \editThree{light} lines.  The photon deficit for a simple occultation are plotted with the \editThree{light, long-dashed lines}.  The excess counts for a zone plate measured over \editTwo{$\twindowshadow$} (\editThree{dark thick line, dotted}) or \editTwo{$\twindowflare$} (\editThree{dark thick line, solid}).  \editThree{Thinner lines with the same line styles are the results for a single lens with chromatic aberration.}\label{fig:NVsR}}
\end{figure*}

\editThree{Of course, maybe ETIs would use smaller lenses, which use less construction material \editTwo{and collide less frequently}.  The signal-to-noise ratio's dependence on $\Rlens$ is shown in Figure~\ref{fig:NVsR} for RXTE-PCA (blue) and HETE-2 FREGATE (black/grey), when all other parameters are fixed.  The minimum radius for $\SNR \ge 5$ with \editTwo{HETE-2 FREGATE} is $\sim 40\ \km$. \editTwo{ETIs might also choose fainter LMXBs as the illumination source, because they are much more common, less mass is needed as ballast against radiation pressure, and the swarm can be more compact while avoiding overheating if collisions can be prevented.  Fainter LMXBs have smaller BL emitting areas.  Under the assumption that the effective temperature of the BL is constant, $\Rstar \propto \sqrt{\Lstar}$.  Then, the peak apparent luminosity of an achromatic lens is invariant but flares are shorter and it is more unlikely for any given lens to transit the NS on any given pass.  As seen in Figure~\ref{fig:NVsR}, $\SNR \propto \Lstar^{1/4}$, so achromatic lenses around faint LMXBs are still detectable throughout the Galaxy.}}

Zone plates \editThree{and lenses with chromatic aberration} with $\bar{E} = 3 k_B \Tstar$ and $\Rlens = 1,000\ \km$ produce far fewer excess photons \editThree{and thus have much lower signal-to-noise.  \emph{Swift}-BAT is particularly insensitive to these events, \editTwo{failing} to detect a lens flare even from the bright, nearby Sco X-1 (2.8 kpc; \citealt{Bradshaw99})\editTwo{, because it can only detect the high energy tail of the LMXB spectrum}.  If $\bar{E}$ is significantly higher (perhaps 10 keV), its prospects should improve.  RHESSI, on the other hand, would have achieved $\SNR > 28$ for Sco X-1 with the same assumptions.  I nonetheless find $\SNR \ge 5$ for all instruments except \emph{Swift}-BAT and sometimes RHESSI, even for lenses with strong chromatic aberration.  \editThree{$\SNR > 10$ is achieved for LMC lens flares for a few sensitive narrow-field instruments like \emph{Chandra} ACIS and \emph{XMM-Newton} EPIC-pnCCD; HETE-2 WXM was marginally sensitive with $\SNR \sim 5$.}  Smaller lenses may very well be missed, as $\SNR \ge 5$ is achieved only for HETE-2 FREGATE when $\Rlens \ga 70\ \km$.  In addition, they would be difficult to detect around the more numerous fainter X-ray binaries.} \editTwo{A luminosity $\ga 10^{37}\ \erg\,\sec^{-1}$ was necessary for HETE-2 FREGATE to achieve $\SNR \sim 5$ for a chromatic lens flare.  Although these events are much dimmer, they may be detected far more often because some photons are re-emitted into a larger cone.  I tested the effects of shifting $b$ and found that HETE-2 FREGATE could detect the fiducial chromatic lens for $b \la 120\ \km$ for the $\twindowshadow$ window, raising the frequency of detectable events by $\ga 10$.  Further gains in $\SNR$ could be achieved by taking advantage of spectral information, as chromatic elements produce an unusual spectral evolution.}

In addition to the rare lens flares, there should be occultation events when the lens passes in front of the BL but magnifies empty space.\footnote{\editThree{The light curve during the transit of a chromatic element is more complex, with obscuration only in a range of energies around $\bar{E}$ and the ``lensed'' spectrum outside of that range (see Figure~\ref{fig:SpectrumMovie}).  In particular shadow events may not even be apparent at $E \gg \bar{E}$ unless $\eta$ is significantly less than $1$.  The following discussion assumes the elements are achromatic lenses or simple occulters.}}   These are much more numerous, roughly by a factor $\sim \Rlens/\Rstar$, but also are much harder to detect, since the photon deficit is bounded by the luminosity of the LMXB and event duration.  From Table~\ref{table:SNR}, \editThree{these events are easily detectable by narrow-field instruments that are observing an occulted LMXB, but the $\SNR$ of a shadow event is low (\editTwo{$\la 10$}) for a wide-field instrument.}  \editThree{Small occulters are detectable by an instrument like RXTE PCA (long-dashed lines in Figure~\ref{fig:NVsR}).  In addition, $1,000\ \km$ radius occulters were detectable around Sco X-1 with \editTwo{$\SNR \sim 70$} with HETE-2, \editTwo{$\sim 20$} for MAXI-GSC, \editTwo{$\sim 16$} for RHESSI, and \editTwo{$\sim 13$} for \emph{Swift}-BAT.}  \editThree{For} the fiducial parameters of Section~\ref{sec:Binary}, the \editThree{occultation} rate is one every $\ga 10\ \Megasec$, and it's conceivable one might be detected during a very \editThree{long} observation of \editThree{Sco X-1 or some other nearby bright LMXB}.

\subsection{Are any known X-ray transients lens flares?}
\label{sec:KnownFlares}
LMXBs are known to flare.  During the common Type I X-ray burst phenomenon, the X-ray luminosity can rise by a factor of a hundred in a fraction of a second, with characteristic temperatures of $2 \endash 3\ \keV$.  Unlike lens flares, though, Type I bursts take tens of seconds to decay, and they are  caused by thermonuclear burning on the NS surface \citep{Strohmayer03,Galloway08}.  Longer types of bursts, like superbursts, also exist and are due to thermonuclear burning \citep{Strohmayer03,intZand19}.  A lens flare would be characterized by a symmetric time profile.  Curiously, though, some bursts have long ($\sim 100\ \sec$) ``eclipse'' like dips, sometimes accompanied by faster ($\sim 1\ \sec$) variability \citep{intZand19}.  It's unlikely occultations would happen only during X-ray bursts, demonstrating that searches for true lens eclipses may have to deal with false positives.

A few X-ray sources display Type II X-ray bursts, classically explained as accretion instabilities.  Type II bursts can have super-Eddington luminosities ($10^{38} \endash 10^{39}\ \erg\,\sec^{-1}$) and can last between a few seconds and a few minutes \citep{Lewin93}.  The longer bursts display a ``flat top'' light curve, with a roughly constant flux followed by a steep decline and a faint tail.  Although vaguely similar to lens flare light curves, long Type II bursts  have a faint tail not expected with lens flares, which break the symmetry of the light curve.  In addition, they last much longer than predicted lens flares; short Type II bursts decay over a prolonged time.  Furthermore, Type II bursts recur frequently, with just a few seconds to a few hours between them \citep{Lewin93}.

Soft gamma repeaters (SGRs), presumed to be magnetars, can emit very powerful X-ray/$\gamma$-ray flares.  These start with an extremely short peak, rising within a millisecond and decaying over a fraction of a second.  They lack the symmetric time curve of lens flare, are much too hot ($\ga 100\ \keV$), and they can be followed by tail emission that decays over minutes (\citealt{Cline80,Hurley05,Palmer05}).  SGRs also emit smaller bursts that do resemble lens flares more: they are cooler ($\sim 9\ \keV$), short ($\sim 0.1\ \sec$), randomly recurring, and some have ``flat top'' light curves with an unresolved rise and fall \citep{Norris91,Thompson95}.  The main arguments against these bursts being lens flares are that 1.) they are somewhat hotter than expected, 2.) the sources are not accreting NSs but magnetars, 3.) the sources emit non-time-symmetric bursts, indicating they naturally flare, and 4.) no long shadow events are observed.

Other types of X-ray transients have been discovered, but they last too long to be lens flares.  Among these are X-ray Flashes, likely off-axis gamma-ray bursts, with durations in the tens of seconds \citep[e.g.,][]{Lamb05}.  Novel extragalactic X-ray transients, including XRT 000519 \citep{Jonker13}, CDF-S XT1 \citep{Bauer17}, and CDF-S XT2 \citep{Xue19} decay slower than they rise and also last for about a minute.  The so-called Fast X-ray Transients are heterogeneous, but last for hours to a day \citep[e.g.,][]{Sguera05}.  Some extragalactic ULXs flare, brightening $\sim 100$ times, but they too last too long and have asymmetric light curves \citep{Sivakoff05,Irwin16}.  

\subsection{Would we have seen lens flares yet?}
Although most X-ray instruments have had narrow fields of view, the \emph{Reuven Ramaty High Energy Solar Spectroscopic Imager} (RHESSI; 2002--2018) could \editThree{have detected} flares from most of the sky \citep{Lin02}.  Thus, a Galactic lens flare \editThree{from a large lens around a bright LMXB} \editThree{in this time frame} would probably have been observed.  The  lack of such flares suggests that either (1) isotropic X-ray lens beacons do not exist around the brighter LMXBs in the Galaxy, (2) the lenses are \editThree{significantly smaller than $1,000\ \km$},  (3) \editThree{the number of lenses in actual swarms is very small, possibly compounded by long orbital periods,} \editThree{or (4) they all have severe chromatic aberration}.

Further constraints on these beacons come from the lack of observed occultations by offset lenses.  Some LMXBs do display partial and total eclipses; the recurrence times are hours, and they probably result from accretion disk structures \citep{Frank87}.  Sco X-1, one of the Galaxy's brightest LMXBs, has been the subject of several searches for millisecond flux drops, which might result from occultations by Kuiper Belt Objects.  After some initial controversy, the occultation rate is now constrained to less than one per $\Megasec$ \citep{Chang16}.  Occultations by misaligned lenses would be much more spectacular \editThree{than Kuiper Belt Object occultations}, lasting several \editThree{seconds} for $\Rlens = 1,000\ \km$.  They should be detectable \editThree{in Sco X-1} even by all-sky monitors.  Reported light curves from these instruments, however, generally bin fluxes into daily measurements \citep[e.g.,][]{McNamara98,Krimm13}.  Some, like \emph{Fermi}-GBM, are non-imaging and cannot distinguish individual nearly-steady sources \citep{Case11}.  \editThree{Although flares should be caught by event triggers,} it is not clear that \editThree{flux decrements from occultations} are adequately constrained yet.  \editThree{RXTE observed Sco X-1 for $1\ \Megasec$ \citep{Chang16}, although the time between eclipses can easily be an order of magnitude longer.  INTEGRAL's total exposure on Sco X-1 is about 4 Ms \citep{Revnivtsev14}.  Because it only measured the flux at 17 keV and above, it could constrain occulters or achromatic lenses, but probably not misaligned elements with chromatic aberration.  \emph{Ariel 5} observed Sco X-1 for five years, during which ten short dips in the flux were observed, but these have been attributed to instrumental effects \citep{Priedhorsky87}.  The \emph{Ariel 5} light curve is binned by 6,000 second orbits \citep{Priedhorsky87}, compared to the $\sim 10\ \sec$ duration of occultations for $\Rlens = 1,000\ \km$, so any actual shadow events would be diluted unless the lenses were much bigger than the Earth.  The Monitor of All-sky X-ray Image (MAXI) observes Sco X-1 for about a minute during each orbit, and after ten years of operation, it has accumulated several Ms of exposure in the relevant energy band.  The effective area is small ($\sim 10\ \cm^2$ for the Gas Slit Camera; \citealt{Sugizaki11}), but sufficient to detect occultations of Sco X-1, and the time resolution is high.  Wide-Field MAXI, which covers 20\% of the sky with a greater effective area \citep{Kawai14}, should provide far more powerful constraints on Sco X-1 occultations.}

\subsection{Would we recognize lens flares as technosignatures?}
\edit1{The large variety of X-ray transients (Section~\ref{sec:KnownFlares}), including those from LMXBs, poses a recognition problem for lens flares.  This is of course a generic problem faced by most proposed technosignatures.  Lens flares have a few characteristics that are not generally expected from X-ray bursts.  Perhaps most importantly, the flares should be accompanied by a larger number of occultation events.  For an achromatic lens, both the rate of occultation events and their duration is proportional to the square root of the peak luminosity enhancement; chromatic elements yield a predictable distribution of spectrum, duration, and peak luminosity for events.  Flares also should occur in the middle of an occultation event.  The shape of the light curve (and possibly spectral variation) also are atypical for X-ray transients, although perhaps these could be explained naturally.  ETIs may engineer secondary technosignatues that are less ambiguous: perhaps instead of having a single $\bar{E}$, they use sub-elements with $\bar{E}$ in ratios of prime numbers, or they use square elements whose shape could be inferred from the light curve. Nonetheless, a candidate event might be explained as some new natural phenomenon, especially since a detection may be too marginal to infer detailed characteristics or to detect occultations.  Follow-up work with a more powerful instrument would be necessary to gain confidence in the artificial nature of the event.}

\edit1{Ambiguous events still allow us to set upper limits on the rate of lens flares, just as we can set upper limits on the prevalence of other technosignatures.  The lack of events with the expected characteristics already allows us constrain these events, demonstrating they are rare or faint in the Milky Way.}

\section{Conclusion}
\label{sec:Conclusion}
\subsection{Summary}
X-ray emitters, particularly neutron stars, are among the most compact long-lived sources of luminosity that exist.  Therefore, they produce the largest flux variations when they are modulated by a structure with a fixed size.  Large X-ray lenses can magnify neutron stars to briefly boost their apparent luminosities by orders of magnitude, a lens flare transient.  In this paper, I considered LMXBs, which give the biggest bang for the buck: a $\Rlens = 1,000\ \km$ achromatic lens can produce flares with luminosities of a billion Suns. 

As a ``beacon'' built by ETIs, lens flares have the advantage of being largely passive.  Like simple occulting structures, they require no electronic systems or power to ``broadcast''.  The sheer brilliance possible with a collimating lens is the greatest allowed by geometrical optics.  The disadvantage of a lens beacon, however, is that the center of \editTwo{an achromatic} lens must pass directly in front of the emitting source to produce a flare.  The effective cross section of the lens for a flare is $\Rstar/\Rlens$ times the geometrical area.  Thus many lenses -- \editThree{about $10^9$} with fiducial parameters -- are required to ensure \editThree{lens flares are frequent enough}.  In turn, the implied mass of the lens system is comparable to a planet; it's not at all clear if enough solid material exists in LMXB systems.  The large number of lenses may prove a guidance problem, if their velocities start to randomize and they begin to crash into one another.  \edit1{Although the lenses do not need to generate vast amounts of power or use complicated beamforming electronics, they could require maintenance to repair damage or deformations that degrade lens performance, particularly due to thermal stress.}

In principle, a flare from an achromatic lens with $\Rlens = 1,000\ \km$ could be detected \editThree{at intergalactic distances} if our most sensitive X-ray instruments was observing the LMXB at the time.  The same detectors could detect a flare from a zone plate of similar size within the \editThree{Galaxy} if it was looking in the right place.  However, lens flares are likely to be extremely rare, occurring perhaps once per decade, and lasting only for about \editThree{$\sim 0.1$} second.  \editThree{Past} all-sky X-ray monitors could detect lens flares within the Milky Way even for $\Rlens \sim 100\ \km$.  These \editThree{have included} \editThree{RHESSI} and FREGATE on HETE-2, which have  operated over most of this century so far.  They \editThree{would have had trouble detecting} the much more common shadow events when a misaligned lens simply blocks the NS instead of magnifying it\editThree{, but dedicated observations with narrow-field instruments can easily detect these}.  Lens flares do not match the properties of observed X-ray bursts of LMXBs.

\subsection{Would it be worth the effort for ETIs?}

Lens flares may be spectacular and attention-grabbing, but would ETIs go to the effort to build a lens system in a hostile, distant environment \edit1{even if they can overcome the many inherent technical problems}?  Our current detector sensitivity limits may be but a passing phase to such a long-lived interstellar society\editTwo{, as emphasized by \citet{Sagan73}.  The Drake equation suggests that ETIs will only be prevalent if their lifetime is very long, and thus their populations may be dominated by societies much older than our own.  On the one hand, the technical capabilities of a typical millennia-old spacefaring society could mitigate the limitations of lens flares.  Over thousands of years, they could accumulate decades of exposure time for all nearby galaxies with X-ray telescopes like our own, or build all-sky X-ray instruments as sensitive as RXTE, for example.  Typical Galactic ETIs might find it easy to detect flares in M31 or further, and conversely, Galactic lens swarm systems may have been built primarily as intergalactic beacons.}

\editTwo{On the other hand, the same technological advancement may eliminate the need for such brilliant events.} It is probably more economical \edit1{and practical} on their end to use much fainter X-ray sources like young neutron stars, or to use occulters\edit1{, small lenses,} or zone plates instead of achromatic lenses \citep[c.f.,][]{Chennamangalam15,Imara18}.  If the typical \editThree{recipient} has capabilities many orders of magnitude beyond our own, LMXBs could be overpowered.  For that matter, there are passive optical beacon systems around normal stars, like the transiting structures described in \citet{Arnold05}.  Although they produce only tiny fluctuations in stellar flux, with our own technology, it is relatively easy to do precision photometry for large fields of stars in the optical.  If most ETIs judge it easier to build optical telescopes able to detect these variations \editTwo{over large distances} than all-sky X-ray monitors somewhat better than ours, they could prefer stellar occulters.  \editTwo{Vast improvements would be necessary to detect stellar occulters over intergalactic distances; we are incapable of detecting a G dwarf in M31, for example.}

It depends on the tradeoffs and intent of the hypothetical ETIs, of course, which we cannot really determine \emph{a priori} -- and likewise, their determination of their audience's capabilities.  The advantage of lenses around LMXBs, however, is that it maximizes the thermal luminosity that can be modulated with kilometer to megameter size structures.  This makes it easier to draw attention over intergalactic distances.  Even if a typical interstellar society has X-ray telescopes a million times more sensitive than ours, for example, using LMXBs could be advantageous because they could be detected over gigaparsec distances.  The use of bright but rare LMXBs could also indicate how much attention the builders want directed, like a monument or advertisement; perhaps they would be preferred locations of ``galactic clubs'' to indicate their reach.  \editTwo{Lens swarms may also be placed around LMXBs for other purposes, namely, beamed power transmission, with the lenses parked in a ``beacon'' constellation in between uses.}

Whatever the case, passive beacons like lens flares are a promising route for SETI.  Although they require a large investment to build, they require \edit1{less} maintenance and might last for geological epochs \edit1{if deformation and collisions can either be repaired or avoided}.  They can also be detected commensally, through the use of wide-field variability observations.  The challenge on our end may simply be patience.

\acknowledgments
{I thank the Breakthrough Listen program for their support. Funding for \emph{Breakthrough Listen} research is sponsored by the Breakthrough Prize Foundation.\footnote{\url{https://breakthroughprize.org/}}  In addition, I acknowledge the use of NASA's Astrophysics Data System and arXiv for this research.  \editThree{I thank the referee\textbf{s} for their comments.}}

\appendix

\section{Orbital stability}
\label{sec:OrbitalStability}
\editThree{There are three problems that arise if the orbits of the lenses evolve rapidly.  First, the lenses may lose their aim towards the barycenter, causing them to magnify only empty space.  Tidal locking may be sufficient to keep the lenses on target if they remain in circular orbits (Appendix~\ref{sec:RotationalStability}), but orbital drift is especially problematic for lenses on eccentric orbits.  For a lens rotation period synchronous with the orbital period, the maximum angular deviation from the system barycenter reaches $\sim 2 e$ for $e \ll 1$.  If the lens has the minimum field of view of $\editTwo{\aNS}/\alens$, an eccentricity as small as $\sim 3 \times 10^{-5} (\editTwo{\aNS}/\Rsun) (\alens/100\ \AU)^{-1}$ will prevent lens flares.  The second reason is that the neutron star will no longer be in focus if the lens drifts too far inwards or outwards, although this only matters for radial migrations of $\ga (\Rstar/\Rlens) \alens$.  Finally, lenses originally placed on non-colliding orbits may wander into each other's paths, initiating a collisional cascade (Appendix~\ref{sec:CollisionalCascades}).}

\editThree{As a measure of the drift of a lens, I use the scale of the neutron star's orbit $\editTwo{\aNS} \sim \Rsun$.  A lens minimal field-of-view starting out in a circular orbit with synchronous rotation will lose its aim if it drifts more than this without some feedback mechanism.  Another relevant scale is the mean separation between the lenses:}
\begin{multline}
s_{LL} \approx \left(\frac{\pi^2}{2} \frac{\Gammaflare}{\alens \aNS \Rstar \Delta_a} \sqrt{\frac{\abin^3}{G \Mbin}}\right)^{1/3} \approx 6\ \Rsun \Delta_a^{1/3} \left(\frac{\Gammaflare}{(30\ \yr)^{-1}}\right)^{-1/3} \left(\frac{\alens}{30\ \AU}\right)^{1/3} \left(\frac{\Rstar}{10\ \km}\right)^{1/3} \left(\frac{\aNS}{\Rsun}\right)^{-1/6} \\
\times \left(\frac{\Mbin}{1.9\ \Msun}\right)^{-1/3} \left(\frac{\Mdonor}{0.5\ \Msun}\right)^{1/2} .
\end{multline}
\editThree{\editTwo{Random d}rifts of this magnitude will generally erase any order that prevents collisions.  In practice, $\editTwo{\aNS}$ is smaller \editTwo{than $s_{LL}$}.}

\subsection{Quadrupole effects from central binary}
\editThree{The gravitational field of the binary deviates from that of a single spherical object, which complicates the dynamics of the swarm.  The lens orbit is not significantly affected by the time-variability of the potential introduced by the binary's mutual orbit, because the binary orbital period is so much shorter than that of the lens.  After averaging over the inner binary's phase for the secular evolution, the potential still has some higher-order terms deviating from a monopole.  The first is a quadrupole term, with magnitude $G \Mstar / \alens \times (\editTwo{\aNS}/\alens)^2$.  If the lens or the binary ever develop highly eccentric orbits, then the octupole term also becomes significant.  This generally should not happen, though, since the binary is very close together and circularized by tidal forces, while the lens would be placed in a circular orbit for attitude maintenance.}

\editThree{A lens and the inner binary form a hierarchical triple system, with the outer component being nearly massless.  The dynamics of these systems has been intensively studied, particularly the Kozai-Lidov mechanism when one member of the inner binary is much smaller than the other two.  \citet{Naoz13} derived the quadrupole and octupole-level evolution of the orbits.  In the relevant quadrupole approximation, the eccentricity $e_O$ and inclination $i_Q$ of the outer orbit remains constant.  Instead, the orbital plane and orbital pericenter both precess.  The ascending node $h_O$ precesses at a rate}
\begin{equation}
|\dot{h_O}| = \frac{-6 C_2 \sin (2 i_{\rm tot})}{L_2 \sqrt{1 - e_O^2}} < \frac{12 C_2}{L_2 \sqrt{1 - e_O^2}}
\end{equation}
\editThree{if the inner eccentricity is zero.  Applying the definitions of $C_2$ and $L_2$ in \citet{Naoz13} to the lens-binary triple, I find}
\begin{equation}
|\dot{h_O}| < \frac{3}{4} \frac{q}{(1 + q)^2} \frac{\sqrt{G \Mbin} \abin^2}{\alens^{7/2}},
\end{equation}
\editThree{where $q$ is the mass-ratio $\MNS/\Mdonor$ of the inner binary.  The drift timescale is then}
\begin{equation}
\tdrift(s) \ga \frac{s}{\alens |\dot{h_0}|} \approx 60\ \kyr \left(\frac{q/(1+q)^2}{3/16}\right) \left(\frac{s}{\Rsun}\right) \left(\frac{\alens}{30\ \AU}\right)^{5/2} \left(\frac{\aNS}{\Rsun}\right)^{-2} \left(\frac{\Mbin}{1.9\ \Msun}\right)^{-5/2} \left(\frac{\Mdonor}{0.5\ \Msun}\right)^{2}.
\end{equation}

\editThree{If the orbit starts out eccentric, its pericenter drifts at a rate}
\begin{equation}
|\dot{g_O}| = 3 C_2 \left[\frac{4 \cos i_{\rm tot}}{G_1} + \frac{10 \cos^2 i_{\rm tot} - 2}{L_2 \sqrt{1 - e_O^2}}\right],
\end{equation}
\editThree{if the inner eccentricity is zero.  From the definition of $G_1$ in \citet{Naoz13}, the $1/G_2$ term is by far more important, so $|\dot{g_O}| < 24 C_2 / G_2$.  The drift time is therefore}
\begin{equation}
\tdrift(s) \ga \frac{s}{\alens |\dot{g_0}|} \approx 30\ \kyr \left(\frac{q/(1+q)^2}{3/16}\right) \left(\frac{s}{\Rsun}\right) \left(\frac{\alens}{30\ \AU}\right)^{5/2} \left(\frac{\aNS}{\Rsun}\right)^{-2} \left(\frac{\Mbin}{1.9\ \Msun}\right)^{-5/2} \left(\frac{\Mdonor}{0.5\ \Msun}\right)^{2}.
\end{equation}
\editThree{Equivalently, the estimated velocity drift is $\sim 0.1\ \cm\,\sec^{-1}$.  This is a conservative estimate, however, since the orbits are likely to be nearly circular.  Furthermore, the octupole equations in \citet{Naoz13} imply the timescale for $e_O$'s evolution is trillions of years.}

\editThree{Thus, it takes many millennia before the quadropole potential terms may start negatively impacting the lens swarm.  \editTwo{Since all the lenses of a given $\alens$ and $i_{\rm tot}$ precess at the same rate, the collision hazard should not be significantly enhanced, and only the aim is affected.} Tidal locking or large fields of view would prolong the useful period of the lens.  Of course, an LMXB differs from the typical detached binary considered in hierarchical triple dynamics, because the system may have mass loss and the donor star is tidally distorted, but those complications should be minor and are beyond the scope of this paper.}

\subsection{Collective gravitational instability in the swarm}
\editThree{The lens swarm might be thought of as a ``gas'' of lenses, at least on smaller scales entirely within the shell.  The swarm could thus experience a gravitational Jeans-like instability.  The timescale is $1/\sqrt{G \nlens \Mlens}$:}
\begin{multline}
\nonumber t_J \approx \left(\frac{\Mbin}{G \abin}\right)^{1/4} \left(\frac{\pi^3 \Rlens^2 \Sigmalens \Gammaflare}{2 \Delta_a \aNS \alens \Rstar}\right)^{-1/2} \approx 58\ \kyr\ \left(\frac{\Mdonor}{0.5\ \Msun}\right)^{3/4} \left(\frac{\Mbin}{1.9\ \Msun}\right)^{-1/2} \left(\frac{\aNS}{\Rsun}\right)^{-1/4} \\
\times \left(\frac{\Rlens}{1,000\ \km}\right)^{-1} \left(\frac{\Sigmalens}{10\ \gcm2}\right)^{-1/2} \left(\frac{\Gammaflare}{(30\ \yr)^{-1}}\right)^{-1/2} \left(\frac{\Delta_a}{0.1}\right)^{1/2} \left(\frac{\alens}{30\ \AU}\right)^{1/2} \left(\frac{\Rstar}{10\ \km}\right)^{1/2} .
\end{multline}
\editThree{In addition, the instability is easily thwarted with a small internal velocity dispersion.  The necessary dispersion is of order $\alens / t_J$:}
\begin{multline}
v_J \approx 2\ \meter\,\sec^{-1} \left(\frac{\alens}{30\ \AU}\right)^{1/2} \left(\frac{\Mdonor}{0.5\ \Msun}\right)^{-3/4} \left(\frac{\Mbin}{1.9\ \Msun}\right)^{1/2} \left(\frac{\aNS}{\Rsun}\right)^{1/4} \\
\times \left(\frac{\Rlens}{1,000\ \km}\right) \left(\frac{\Sigmalens}{10\ \gcm2}\right)^{1/2} \left(\frac{\Gammaflare}{(30\ \yr)^{-1}}\right)^{1/2} \left(\frac{\Delta_a}{0.1}\right)^{-1/2} \left(\frac{\Rstar}{10\ \km}\right)^{-1/2} .
\end{multline}

\subsection{Effect of lens-lens encounters}
\editThree{A final effect I consider here is individual lens-lens interactions.  When lenses pass each other without colliding, they still exert a gravitational force on each other.  Over time, the cumulative effect of these encounters will both induce drift in the lenses and increase their velocity dispersion.}

\editThree{I treat the effect of a lens-lens encounter using an impulse approximation, in which the encounter's overall effect is a small impulse in velocity towards the other lens at the point of closest approach.  An incoming lens projectile with velocity $v$ and impact parameter $b$ induces a velocity impulse of}
\begin{equation}
\Delta v \approx \frac{2 G \Mlens}{b \vrand} .
\end{equation}
\editThree{Over a time $t$, the lens experiences many such encounters at a variety of distances.  Because these are random encounters, the total velocity change from these encounters is roughly a Gaussian with a variance equal to the sum of the squares of the velocity impulses:}
\begin{equation}
\sigma_v^2 (t) \approx \int_{b_{\rm min}}^{b_{\rm max}} t \frac{d\Gamma}{db} \left(\frac{2 G \Mlens}{b \vrand}\right)^2 db,
\end{equation}
\editThree{where $d\Gamma/db = 2 \pi \nlens b \vrand$ is the rate of encounters with impact parameter $b$.  The minimum impact parameter is given by the mean closest $b$ over an interval $t$, and is roughly $1/\sqrt{\pi \nlens t \vrand}$.  The maximum impact parameter is set by the requirement that an encounter is shorter than the mean time between encounters, and is thus $1/(4 \pi \nlens)^{1/3}$, of order the mean distance between lenses.  The time to heat the lens population by $\sigma_v$ is}
\begin{equation}
t_v \approx \frac{\sigma_v^2 \vrand}{8 \pi G^2 \nlens \Mlens^2 \ln \Lambda} \approx \sigma_v^2 \frac{\vrand}{\vlens} \sqrt{\frac{(1-\beta) \alens}{\abin^3}} \frac{\Delta_a \Mbin \aNS \Rstar}{4 \pi^5 G \Sigmalens^2 \Rlens^4 \Gammaflare \ln \Lambda} ,
\end{equation}
\editThree{with $\Lambda \equiv b_{\rm max}/b_{\rm min}$.  This time is nearly a quadrillion years even for $\sigma_v = 1\ \meter\,\sec^{-1}$.}  

\editThree{But even the small velocity jolts might add up over geological epochs.  Each encounter produces a mean displacement of $\Delta v t / 2$ over a time $t$, since the typical encounter occurs halfway through the interval.  The total displacement is also roughly a Gaussian with a variance:}
\begin{equation}
\sigma_s^2 (t) \approx \int_{b_{\rm min}}^{b_{\rm max}} \frac{t^3}{4} \frac{d\Gamma}{db} \left(\frac{2 G \Mlens}{b v}\right)^2 db .
\end{equation}
\editThree{Finally, the drift time is}
\begin{multline}
\tdrift \approx \left[\sigma_s^2 \frac{\vrand}{\vlens} \sqrt{\frac{(1-\beta) \alens}{\abin^3}} \frac{\Delta_a \Mbin \aNS \Rstar}{\pi^5 G \Sigmalens^2 \Rlens^4 \Gammaflare \ln \Lambda}\right]^{1/3} \approx 0.66\ \Myr\ (1 - \beta)^{1/6} \left(\frac{\sigma_s}{\Rsun}\right)^{2/3}  \left(\frac{\vrand}{\vlens}\right)^{1/3} \left(\frac{\alens}{30\ \AU}\right)^{1/6} \\
\times \left(\frac{\aNS}{\Rsun}\right)^{-1/6}   \left(\frac{\Mbin}{1.9\ \Msun}\right)^{-1/6} \left(\frac{\Mdonor}{0.5\ \Msun}\right)^{1/2} \left(\frac{\Delta_a}{0.1}\right)^{1/3} \left(\frac{\Rstar}{10\ \km}\right)^{1/3} \left(\frac{\Sigmalens}{10\ \gcm2}\right)^{-2/3} \left(\frac{\Rlens}{1,000\ \km}\right)^{-4/3} \\
\times \left(\frac{\Gammaflare}{(30\ \yr)^{-1}}\right)^{-1/3} \left(\frac{\ln \Lambda}{10}\right)^{-1/3} .
\end{multline}
\editThree{The lens swarm should therefore be safe from this effect for hundreds of millennia.}

\section{Collisional cascades}
\label{sec:CollisionalCascades}
\editThree{The initial mean collisional rate is low enough that the swarm can be expected to avoid destruction for millennia.  Unfortunately, direct collisions with other intact lenses is not the only threat.  Each collision produces a spray of debris, with several pieces themselves large enough to destroy a lens.  This is the analog of Kessler syndrome, a collisional cascade that operates among Earth satellites \citep{Kessler78}.}  

\editThree{If placed on truly random orbits, the first lens collision would be expected within a year, initiating the cascade.  A very rough lower estimate of the time it takes for the cascade to develop is found by assuming that each collision produces $D$ large pieces of debris, each large enough to destroy another orbiting object.  Here, I neglect the size differences between debris and the lenses, treating them all as a single population of $N$ objects.  Then the population evolves as $dN/dt \approx D N^2 A \vrand / V$, where $V$ is the volume occupied by the objects.  Then $N(t) = (1/N(0) - D A \vrand t / V)^{-1}$.  The cascade becomes overwhelming on a timescale $(D n \vrand A)^{-1}$, faster by $D$ than the non-cascade expectation.  In actuality, much of the debris from second-generation collisions and later is too small to destroy a lens.}

\editThree{For Earth satellites, \citet{Kessler78} estimate the number of pieces of debris with mass $\ge m$ produced by a collision at $10\ \kms$ as $0.4 (m / M_{\rm ej})^{-0.8}$, where $M_{\rm ej}$ is the mass ejected by the collision.  They also estimate a collision is catastrophic if the mass ratio is $\la 1,000$.  Thus each collision between satellites should produce $\sim 175$ additional pieces of debris.  Collisions with mass ratios $\sim 10^5$ may be expected to produce at least one piece of debris large enough to catastrophically damage a satellite.  A very rough, conservative estimate of the cascade development time is therefore $\sim 1/200$ of the naive collision time.}

\editThree{Of course, it is a risky endeavor to apply these estimates to these lenses.  It is unclear how the thin structure of the assemblage would affect the estimates, and for all we know, they might be made of superstrong materials.  In addition, the lenses may be initially started out with low relative velocities to each other, reducing both the rate of collisions and their damage.  It is also unclear how much sub-catastrophic collisions would degrade lens performance, as they punch holes in the lens and its support structures.}

\editThree{The comparatively large luminosity from LMXBs serves as an advantage in this case, though.  Small enough pieces of debris are blown out of the system entirely by radiation pressure (Section~\ref{sec:Luminosity}).  This is a rapid process, taking only}
\begin{equation}
t_{\rm rad} \approx \sqrt{\frac{4 \pi \Delta \alens^3 c \Sigma}{\zetaabs \Lstar}} \approx 36\ \yr \left(\frac{\alens}{100\ \AU}\right)^{3/2} \left(\frac{\zetaabs \Sigma}{\gcm2}\right)^{1/2} \left(\frac{\Lstar}{10^{38}\ \erg\ \sec^{-1}}\right)^{-1/2} \left(\frac{\Delta}{0.1}\right)^{1/2}.
\end{equation}
\editThree{LMXB lens swarms may be self-cleaning.}

\section{Rotational stability}
\label{sec:RotationalStability}
\editThree{A plate orbiting with its face normal to a central object is rotationally unstable.  The tides of the central object will excite any small perturbation away from being face-on.  Consider a solid cylinder of radius $r$, height $h$, and mass $m$.  Let $\hat{Z}$ be the direction lengthwise along the cylinder, with $\hat{X}$ and $\hat{Y}$ being normal to the sides of the cylinder and each other.  Now consider the gravitational tides from a central source of mass $M$ located in the direction $\hat{r} = \sin \phi \hat{X} + \cos \phi \hat{Z}$ at a distance of $R \gg \max(h, r)$.  The tidal force on each mass element in the cylinder is:}
\begin{equation}
\frac{d\vec{F}_{\rm tidal}}{dm} = \frac{2 G M}{R^3} (X \sin \phi + Z \cos \phi) (\sin \phi \hat{X} + \cos \phi \hat{Z}),
\end{equation}
\editThree{resulting in a torque per unit volume of:}
\begin{equation}
\frac{d\vec{\tau}}{dm} = \frac{2 G M}{R^3} (X \sin \phi + Z \cos \phi) (Y \cos \phi \hat{X} + (Z \sin \phi - X \cos \phi) \hat{Y} -  Y \sin \phi \hat{Z}).
\end{equation}
\editThree{Integrating over the volume of the cylinder results in a net torque of} 
\begin{equation}
\tau_Y = \frac{\pi G M m}{R^3} \sin 2\phi \left(\frac{h^2}{12} - \frac{r^2}{4}\right)
\end{equation}
\editThree{in the $\hat{Y}$ direction.  The behavior of $\phi$ then can be estimated using the moment of inertia along the $\hat{Y}$ axis through the cylinder's center of mass, $m/12 (3 r^2 + h^2)$.  A positive $\tau_Y$ results in the cylinder being torqued so its end faces point towards the central object, whereas a negative $\tau_Y$ results in its rim being torqued to face it.  To maintain the pointing of a lens so it faces a star, a positive $\tau_Y$ is needed, but because the lens is thin ($r \gg h$), this attitude is unstable.  This instability is rapid, with a growth timescale of $P/(2\pi) \times \sqrt{(3 r^2 + h^2)/(3r^2 - h^2)}$, where $P$ is the period of the orbit of a cylinder in circular orbit at $R$ from the central object.}

\editThree{This problem can be addressed by attaching a counterbalance to the lens that juts out towards and away from the star.  The simplest would be to place a long, thin ``pole'' through the center of the lens.  In order to cancel out the tidal torque on the lens disk, the pole needs a radius of $R_{\rm pole} = \sqrt{3 \Sigmalens \Rlens^4 / (\rho_{\rm pole} h_{\rm pole}^3)}$.  For a pole made of carbon ($\rho_{\rm pole} = 2\ \rhoUnits$) and with $h_{\rm pole} \approx \Rlens$, the ``pole'' needs to have a radius of:}
\begin{equation}
R_{\rm pole} > 0.39\ \km \left(\frac{\Sigmalens}{\gcm2}\right)^{1/2} \left(\frac{\rho_{\rm pole}}{2\ \rhoUnits}\right)^{-1/2} \left(\frac{h_{\rm pole}}{\Rlens}\right)^{-3/2} \left(\frac{\Rlens}{1,000\ \km}\right)^{-1/2}.
\end{equation}
\editThree{The actual radius may need to be substantially larger to robustly prevent rotation.  A solid pole is not necessary, however.  One can imagine using a material like carbon nanotubes, proposed for use in a space elevator, to construct a dense ``forest'' of long tethers.  Alternatively, a squat ``pole'' of radius $\sim 10\ \km$ and length $\sim 100\ \km$ could also work.  The mass in the pole is $3 (\Rlens / h_{\rm pole})^2 M_{\rm disk}$.  Unless the pole is extremely long, it will dominate the mass of the artifact.  However, the additional mass of the pole can serve a second use as ``ballast'' when the central source is very bright, to prevent the lens from being blown away by radiation pressure.  A disadvantage of using a pole is that it could reduce efficiency by introducing sidelobes into the lens output.  Another option is to enclose the lens in a tube of length $>\sqrt{6} \Rlens$.  A tube would necessarily restrict the field of view to $<2/\sqrt{6}$ radians, although this should not be a problem as long as eccentricity is low.}

\editThree{Without a form of dissipation, a perturbation will result in the lens-counterweight system oscillating without end.  If dissipation can be provided in the assemblage, however, it will become tidally locked to the barycenter of the system.  Dissipation might be accomplished by using reservoirs of ball bearings that slosh around and convert kinetic energy into heat.  This is advantageous for a circular orbit, as it results in the X-ray source remaining in the field of view of the lens.  The orbit cannot develop too high an eccentricity, or the rotation may assume some other ratio than 1:1 compared to the orbital period.}



\begin{thebibliography}{}
\twocolumngrid

\bibitem[Alpar et al.(1982)]{Alpar82} Alpar, M.~A., Cheng, A.~F., Ruderman, M.~A., et al.\ 1982, \nat, 300, 728

\bibitem[Arnold(2005)]{Arnold05} Arnold, L.~F.~A.\ 2005, \apj, 627, 534

\bibitem[Arzoumanian et al.(2014)]{Arzoumanian14} Arzoumanian, Z., Gendreau, K.~C., Baker, C.~L., et al.\ 2014, \procspie, 914420

\bibitem[Atteia et al.(2003)]{Atteia03} Atteia, J.-L., Boer, M., Cotin, F., et al.\ 2003, Gamma-ray Burst and Afterglow Astronomy 2001: A Workshop Celebrating the First Year of the HETE Mission, 17

\bibitem[Bachetti et al.(2014)]{Bachetti14} Bachetti, M., Harrison, F.~A., Walton, D.~J., et al.\ 2014, \nat, 514, 202

\bibitem[Banit et al.(1993)]{Banit93} Banit, M., Ruderman, M.~A., Shaham, J., et al.\ 1993, \apj, 415, 779

\bibitem[Barcons et al.(2015)]{Barcons15} Barcons, X., Nandra, K., Barret, D., et al.\ 2015, Journal of Physics Conference Series, 012008

\bibitem[Bauer et al.(2017)]{Bauer17} Bauer, F.~E., Treister, E., Schawinski, K., et al.\ 2017, \mnras, 467, 4841

\bibitem[Baumgartner et al.(2013)]{Baumgartner13} \editThree{Baumgartner, W.~H., Tueller, J., Markwardt, C.~B., et al.\ 2013, \apjs, 207, 19}

\bibitem[Begelman et al.(2006)]{Begelman06} Begelman, M.~C., King, A.~R., \& Pringle, J.~E.\ 2006, \mnras, 370, 399

\bibitem[Bildsten \& Deloye(2004)]{Bildsten04} Bildsten, L., \& Deloye, C.~J.\ 2004, \apjl, 607, L119 

\bibitem[Boe et al.(2019)]{Boe19} \editTwo{Boe, B., Jedicke, R., Meech, K.~J., et al.\ 2019, \icarus, 333, 252, arXiv:1905.13458}

\bibitem[Bracewell(1960)]{Bracewell60} Bracewell, R.~N.\ 1960, \nat, 186, 670

\bibitem[Bradshaw et al.(1999)]{Bradshaw99} Bradshaw, C.~F., Fomalont, E.~B., \& Geldzahler, B.~J.\ 1999, \apjl, 512, L121

\bibitem[Bradt et al.(1993)]{Bradt93} \editTwo{Bradt, H.~V., Rothschild, R.~E., \& Swank, J.~H.\ 1993, \aaps, 97, 355}

\bibitem[Braig \& Predehl(2012)]{Braig12} \editThree{Braig, C., \& Predehl, P.\ 2012, \ao, 51, 4638}

\bibitem[Brook et al.(2014)]{Brook14} Brook, P.~R., Karastergiou, A., Buchner, S., et al.\ 2014, \apjl, 780, L31

\bibitem[Campana et al.(2011)]{Campana11} Campana, S., Lodato, G., D'Avanzo, P., et al.\ 2011, \nat, 480, 69

\bibitem[Carrigan(2009)]{Carrigan09} Carrigan, R.~A., Jr.\ 2009, \apj, 698, 2075 

\bibitem[Carstairs(2002)]{Carstairs02} Carstairs, I.~R.\ 2002, Astronomy and Geophysics, 43, 6.26 

\bibitem[Case et al.(2011)]{Case11} Case, G.~L., Cherry, M.~L., Wilson-Hodge, C.~A., et al.\ 2011, \apj, 729, 105 

\bibitem[Caves, \& Drummond(1994)]{Caves94} Caves, C.~M., \& Drummond, P.~D.\ 1994, Reviews of Modern Physics, 66, 481

\bibitem[Chang et al.(2016)]{Chang16} Chang, H.-K., Liu, C.-Y., \& Shang, J.-R.\ 2016, \mnras, 462, 1952 

\bibitem[Chennamangalam et al.(2015)]{Chennamangalam15} Chennamangalam, J., Siemion, A.~P.~V., Lorimer, D.~R., \& Werthimer, D.\ 2015, \na, 34, 245 

\bibitem[Cline et al.(1980)]{Cline80} Cline, T.~L., Desai, U.~D., Pizzichini, G., et al.\ 1980, \apjl, 237, L1

\bibitem[Corbet(1997)]{Corbet97} Corbet, R.~H.~D.\ 1997, Journal of the British Interplanetary Society, 50, 253, arXiv:1609.00330

\bibitem[Cordes \& Shannon(2008)]{Cordes08} Cordes, J.~M., \& Shannon, R.~M.\ 2008, \apj, 682, 1152

\bibitem[Currie \& Hansen(2007)]{Currie07} Currie, T., \& Hansen, B.\ 2007, \apj, 666, 1232

\bibitem[Di Stefano \& Ray(2016)]{diStefano16} Di Stefano, R. \& Ray, A.\ 2016, \apj, 827, 54

\bibitem[Fabbiano(2006)]{Fabbiano06} Fabbiano, G.\ 2006, \araa, 44, 323 

\bibitem[Feroci et al.(2007)]{Feroci07} Feroci, M., Costa, E., Soffitta, P., et al.\ 2007, Nuclear Instruments and Methods in Physics Research A, 581, 728

\bibitem[Forgan \& Nichol(2011)]{Forgan11} Forgan, D.~H., \& Nichol, R.~C.\ 2011, International Journal of Astrobiology, 10, 77

\bibitem[Frank et al.(1987)]{Frank87} Frank, J., King, A.~R., \& Lasota, J.-P.\ 1987, \aap, 178, 137

\bibitem[Galloway et al.(2008)]{Galloway08} Galloway, D.~K., Muno, M.~P., Hartman, J.~M., Psaltis, D., \& Chakrabarty, D.\ 2008, \apjs, 179, 360 

\bibitem[Garmire et al.(2003)]{Garmire03} \editThree{Garmire, G.~P., Bautz, M.~W., Ford, P.~G., et al.\ 2003, \procspie, 28}

\bibitem[Gendreau et al.(2012)]{Gendreau12} Gendreau, K.~C., Arzoumanian, Z., \& Okajima, T.\ 2012, \procspie, 844313

\bibitem[Geng \& Huang(2015)]{Geng15} Geng, J.~J., \& Huang, Y.~F.\ 2015, \apj, 809, 24

\bibitem[Gilfanov et al.(2003)]{Gilfanov03} Gilfanov, M., Revnivtsev, M., \& Molkov, S.\ 2003, \aap, 410, 217 

\bibitem[Gilfanov(2004)]{Gilfanov04} Gilfanov, M.\ 2004, \mnras, 349, 146 

\bibitem[Grimm et al.(2002)]{Grimm02} Grimm, H.-J., Gilfanov, M., \& Sunyaev, R.\ 2002, \aap, 391, 923 

\bibitem[Hankins et al.(2003)]{Hankins03} Hankins, T.~H., Kern, J.~S., Weatherall, J.~C., \& Eilek, J.~A.\ 2003, \nat, 422, 141 

\bibitem[Hankins \& Eilek(2007)]{Hankins07} Hankins, T.~H., \& Eilek, J.~A.\ 2007, \apj, 670, 693 

\bibitem[Hansen et al.(2009)]{Hansen09} Hansen, B.~M.~S., Shih, H.-Y., \& Currie, T.\ 2009, \apj, 691, 382

\bibitem[Harrison et al.(2013)]{Harrison13} Harrison, F.~A., Craig, W.~W., Christensen, F.~E., et al.\ 2013, \apj, 770, 103

\bibitem[Hippke \& Forgan(2017a)]{Hippke17-OptimalNu} Hippke, M., \& Forgan, D.~H.\ 2017a, arXiv:1711.05761 

\bibitem[Hippke \& Forgan(2017b)]{Hippke17-XSearch} Hippke, M., \& Forgan, D.~H.\ 2017b, arXiv:1712.06639

\bibitem[Hippke(2018)]{Hippke18-MaxInfo} Hippke, M.\ 2018, arXiv:1801.06218

\bibitem[Huang \& Geng(2014)]{Huang14} Huang, Y.~F., \& Geng, J.~J.\ 2014, \apjl, 782, L20

\bibitem[Hubbell \& Seltzer(1996)]{Hubbell96} Hubbell, J. H., \& Seltzer, S. M.\ 1996, ``NIST standard reference database 126: X-Ray Mass Attenuation Coefficients'', Gaithersburg, MD: National Institute of Standards and Technology, https://www.nist.gov/pml/x-ray-mass-attenuation-coefficients

\bibitem[Hurley et al.(2005)]{Hurley05} Hurley, K., Boggs, S.~E., Smith, D.~M., et al.\ 2005, \nat, 434, 1098

\bibitem[Imara \& Di Stefano(2018)]{Imara18} Imara, N., \& Di Stefano, R.\ 2018, \apj, 859, 40

\bibitem[Inogamov \& Sunyaev(1999)]{Inogamov99} Inogamov, N.~A., \& Sunyaev, R.~A.\ 1999, Astronomy Letters, 25, 269 

\bibitem[in't Zand et al.(2019)]{intZand19} in't Zand, J.~J.~M., Kries, M.~J.~W., Palmer, D.~M., et al.\ 2019, \aap, 621, A53

\bibitem[Irwin et al.(2016)]{Irwin16} Irwin, J.~A., Maksym, W.~P., Sivakoff, G.~R., et al.\ 2016, \nat, 538, 356

\bibitem[Israel et al.(2017)]{Israel17} Israel, G.~L., Belfiore, A., Stella, L., et al.\ 2017, Science, 355, 817

\bibitem[Jahoda et al.(1996)]{Jahoda96} Jahoda, K., Swank, J.~H., Giles, A.~B., et al.\ 1996, \procspie, 59

\bibitem[Jonker et al.(2013)]{Jonker13} Jonker, P.~G., Glennie, A., Heida, M., et al.\ 2013, \apj, 779, 14

\bibitem[Karachentsev et al.(2002)]{Karachentsev02} \editTwo{Karachentsev, I.~D., Dolphin, A.~E., Geisler, D., et al.\ 2002, \aap, 383, 125}

\bibitem[Kawai et al.(2014)]{Kawai14} \editThree{Kawai, N., Tomida, H., Yatsu, Y., et al.\ 2014, \procspie, 91442P}

\bibitem[Kerr et al.(2015)]{Kerr15} Kerr, M., Johnston, S., Hobbs, G., \& Shannon, R.~M.\ 2015, \apjl, 809, L11 

\bibitem[Kessler \& Cour-Palais(1978)]{Kessler78} \editThree{Kessler, D.~J., \& Cour-Palais, B.~G.\ 1978, \jgr, 83, 2637}

\bibitem[Kipping(2019)]{Kipping19} Kipping, D.\ 2019, Research Notes of the American Astronomical Society, 3, 91

\bibitem[Krimm et al.(2013)]{Krimm13} Krimm, H.~A., Holland, S.~T., Corbet, R.~H.~D., et al.\ 2013, \apjs, 209, 14

\bibitem[Lacki(2016)]{Lacki16-K3} Lacki, B.~C.\ 2016, arXiv:1604.07844 

\bibitem[Lacki(2019)]{Lacki19-Glint} Lacki, B.~C.\ 2019, \pasp, 131, 084401 

\bibitem[Lamb et al.(2005)]{Lamb05} Lamb, D.~Q., Donaghy, T.~Q., \& Graziani, C.\ 2005, \apj, 620, 355

\bibitem[Lewin et al.(1993)]{Lewin93} Lewin, W.~H.~G., van Paradijs, J., \& Taam, R.~E.\ 1993, \ssr, 62, 223 

\bibitem[Lin et al.(1991)]{Lin91} Lin, D.~N.~C., Woosley, S.~E., \& Bodenheimer, P.~H.\ 1991, \nat, 353, 827

\bibitem[Lin et al.(2002)]{Lin02} Lin, R.~P., Dennis, B.~R., Hurford, G.~J., et al.\ 2002, \solphys, 210, 3

\bibitem[Linares et al.(2010)]{Linares10} Linares, M., Watts, A., Altamirano, D., et al.\ 2010, \apjl, 719, L84

\bibitem[Lorimer et al.(2007)]{Lorimer07} Lorimer, D.~R., Bailes, M., McLaughlin, M.~A., Narkevic, D.~J., \& Crawford, F.\ 2007, Science, 318, 777 

\bibitem[Margalit \& Metzger(2017)]{Margalit17} Margalit, B., \& Metzger, B.~D.\ 2017, \mnras, 465, 2790

\bibitem[Martin et al.(2016)]{Martin16} Martin, R.~G., Livio, M., \& Palaniswamy, D.\ 2016, \apj, 832, 122 

\bibitem[Mata S\'anchez et al.(2015)]{MataSanchez15} Mata S\'anchez, D., Munoz-Darias, T., Casares, J., et al.\ 2015, \mnras, 449, L1

\bibitem[Matsuoka et al.(2009)]{Matsuoka09} \editThree{Matsuoka, M., Kawasaki, K., Ueno, S., et al.\ 2009, \pasj, 61, 999}

\bibitem[McConnachie et al.(2005)]{McConnachie05} \editThree{McConnachie, A.~W., Irwin, M.~J., Ferguson, A.~M.~N., et al.\ 2005, \mnras, 356, 979}

\bibitem[McNamara et al.(1998)]{McNamara98} \editThree{McNamara, B.~J., Harrison, T.~E., Mason, P.~A., et al.\ 1998, \apjs, 116, 287}

\bibitem[Merloni et al.(2012)]{Merloni12} Merloni, A., Predehl, P., Becker, W., et al.\ 2012, arXiv e-prints, arXiv:1209.3114

\bibitem[Miller \& Hamilton(2001)]{Miller01} Miller, M.~C., \& Hamilton, D.~P.\ 2001, \apj, 550, 863 

\bibitem[Mitsuda et al.(1984)]{Mitsuda84} \editThree{Mitsuda, K., Inoue, H., Koyama, K., et al.\ 1984, \pasj, 36, 741}

\bibitem[Mottez et al.(2013)]{Mottez13} Mottez, F., Bonazzola, S., \& Heyvaerts, J.\ 2013, \aap, 555, A126

\bibitem[Naoz et al.(2013)]{Naoz13} \editThree{Naoz, S., Farr, W.~M., Lithwick, Y., et al.\ 2013, \mnras, 431, 2155}

\bibitem[Norris et al.(1991)]{Norris91} Norris, J.~P., Hertz, P., Wood, K.~S., et al.\ 1991, \apj, 366, 240

\bibitem[Osmanov(2016)]{Osmanov16} Osmanov, Z.\ 2016, International Journal of Astrobiology, 15, 127 

\bibitem[Palmer et al.(2005)]{Palmer05} Palmer, D.~M., Barthelmy, S., Gehrels, N., et al.\ 2005, \nat, 434, 1107

\bibitem[Pavlov et al.(2002)]{Pavlov02} Pavlov, G.~G., Zavlin, V.~E., \& Sanwal, D.\ 2002, Neutron Stars, Pulsars, and Supernova Remnants, 273

\bibitem[Pavlov et al.(2009)]{Pavlov09} Pavlov, G.~G., Kargaltsev, O., Wong, J.~A., et al.\ 2009, \apj, 691, 458

\bibitem[Pfahl et al.(2003)]{Pfahl03} Pfahl, E., Rappaport, S., \& Podsiadlowski, P.\ 2003, \apj, 597, 1036 

\bibitem[Phinney \& Hansen(1993)]{Phinney93} Phinney, E.~S., \& Hansen, B.~M.~S.\ 1993, Planets Around Pulsars, 371

\bibitem[Pietrzy{\'n}ski et al.(2019)]{Pietrzynski19} \editTwo{Pietrzy{\'n}ski, G., Graczyk, D., Gallenne, A., et al.\ 2019, \nat, 567, 200}

\bibitem[Podsiadlowski(1993)]{Podsiadlowski93} Podsiadlowski, P.\ 1993, Planets Around Pulsars, 149

\bibitem[Popham \& Sunyaev(2001)]{Popham01} Popham, R., \& Sunyaev, R.\ 2001, \apj, 547, 355 

\bibitem[Posselt et al.(2012)]{Posselt12} Posselt, B., Pavlov, G.~G., Manchester, R.~N., et al.\ 2012, \apj, 749, 146

\bibitem[Priedhorsky \& Holt(1987)]{Priedhorsky87} \editThree{Priedhorsky, W.~C., \& Holt, S.~S.\ 1987, \apj, 312, 743}

\bibitem[Rasio et al.(1992)]{Rasio92} Rasio, F.~A., Shapiro, S.~L., \& Teukolsky, S.~A.\ 1992, \aap, 256, L35

\bibitem[Rau et al.(2013)]{Rau13} \editThree{Rau, A., Meidinger, N., Nandra, K., et al.\ 2013, arXiv e-prints, arXiv:1308.6785}

\bibitem[Revnivtsev \& Gilfanov(2006)]{Revnivtsev06} Revnivtsev, M.~G., \& Gilfanov, M.~R.\ 2006, \aap, 453, 253

\bibitem[Revnivtsev et al.(2013)]{Revnivtsev13} Revnivtsev, M.~G., Suleimanov, V.~F., \& Poutanen, J.\ 2013, \mnras, 434, 2355 

\bibitem[Revnivtsev et al.(2014)]{Revnivtsev14} \editThree{Revnivtsev, M.~G., Tsygankov, S.~S., Churazov, E.~M., et al.\ 2014, \mnras, 445, 1205}

\bibitem[Rybicki \& Lightman(1979)]{Rybicki79} Rybicki, G.~B., \& Lightman, A.~P.\ 1979, New York, Wiley-Interscience  

\bibitem[Sagan(1973)]{Sagan73} Sagan, C.\ 1973, \icarus, 19, 350 

\bibitem[Sallmen et al.(2019)]{Sallmen19} Sallmen, S., Korpela, E.~J., \& Crawford-Taylor, K.\ 2019, \aj, 158, 258

\bibitem[Schwartz(2014)]{Schwartz14} Schwartz, D.~A.\ 2014, Review of Scientific Instruments, 85, 061101

\bibitem[Sguera et al.(2005)]{Sguera05} Sguera, V., Barlow, E.~J., Bird, A.~J., et al.\ 2005, \aap, 444, 221

\bibitem[Shannon et al.(2013)]{Shannon13} Shannon, R.~M., Cordes, J.~M., Metcalfe, T.~S., et al.\ 2013, \apj, 766, 5

\bibitem[Shirasaki et al.(2003)]{Shirasaki03} Shirasaki, Y., Kawai, N., Yoshida, A., et al.\ 2003, \pasj, 55, 1033

\bibitem[Sivakoff et al.(2005)]{Sivakoff05} Sivakoff, G.~R., Sarazin, C.~L., \& Jord{\'a}n, A.\ 2005, \apjl, 624, L17

\bibitem[Skinner(2001)]{Skinner01} Skinner, G.~K.\ 2001, \aap, 375, 691

\bibitem[Skinner(2002)]{Skinner02} Skinner, G.~K.\ 2002, \aap, 383, 352 

\bibitem[Skinner(2010)]{Skinner10} Skinner, G.~K.\ 2010, X-Ray Optics and Instrumentation, 2010, 743485

\bibitem[Smith et al.(2002)]{Smith02} \editThree{Smith, D.~M., Lin, R.~P., Turin, P., et al.\ 2002, \solphys, 210, 33}

\bibitem[Stevens et al.(1992)]{Stevens92} Stevens, I.~R., Rees, M.~J., \& Podsiadlowski, P.\ 1992, \mnras, 254, 19P

\bibitem[Strohmayer \& Bildsten(2003)]{Strohmayer03} Strohmayer, T., \& Bildsten, L.\ 2003, arXiv e-prints, astro-ph/0301544

\bibitem[Str{\"u}der et al.(2001)]{Struder01} \editThree{Str{\"u}der, L., Briel, U., Dennerl, K., et al.\ 2001, \aap, 365, L18}

\bibitem[Sugizaki et al.(2011)]{Sugizaki11} \editThree{Sugizaki, M., Mihara, T., Serino, M., et al.\ 2011, \pasj, 63, S635}

\bibitem[Sunyaev, \& Shakura(1986)]{Sunyaev86} Sunyaev, R.~A., \& Shakura, N.~I.\ 1986, Soviet Astronomy Letters, 12, 117

\bibitem[Swift-BAT Team(2005)]{SwiftBAT05} Swift-BAT Team 2005, ``BAT Digest'', https://swift.gsfc.nasa.gov/analysis/bat\_digest.html

\bibitem[Tanabashi et al.(2018)]{Tanabashi18} Tanabashi, M., Hagiwara, K., Hikasa, K., et al.\ 2018, \prd, 98, 030001

\bibitem[Tavani \& Brookshaw(1992)]{Tavani92} Tavani, M., \& Brookshaw, L.\ 1992, \nat, 356, 320

\bibitem[Thompson, \& Duncan(1995)]{Thompson95} Thompson, C., \& Duncan, R.~C.\ 1995, \mnras, 275, 255

\bibitem[van der Klis(1989)]{vanDerKlis89} van der Klis, M.\ 1989, \araa, 27, 517

\bibitem[Wang et al.(2003)]{Wang03} Wang, Y., Yun, W., \& Jacobsen, C.\ 2003, \nat, 424, 50

\bibitem[Wang et al.(2014)]{Wang14} Wang, Z., Ng, C.-Y., Wang, X., et al.\ 2014, \apj, 793, 89

\bibitem[Webbink et al.(1983)]{Webbink83} Webbink, R.~F., Rappaport, S., \& Savonije, G.~J.\ 1983, \apj, 270, 678 

\bibitem[Weisskopf et al.(2003)]{Weisskopf03} \editTwo{Weisskopf, M.~C., Aldcroft, T.~L., Bautz, M., et al.\ 2003, Experimental Astronomy, 16, 1}

\bibitem[Weissman(1996)]{Weissman96} Weissman, P.~R.\ 1996, Completing the Inventory of the Solar System, 265

\bibitem[Wolszczan \& Frail(1992)]{Wolszczan92} Wolszczan, A., \& Frail, D.~A.\ 1992, \nat, 355, 145

\bibitem[Xue et al.(2019)]{Xue19} Xue, Y.~Q., Zheng, X.~C., Li, Y., et al.\ 2019, \nat, 568, 198

\bibitem[XMM-Newton SOC(2018)]{XMMNewtonSOC18} XMM-Newton Science Operations Center, ``XMM-Newton Users Handbook'', Issue 2.16, 2018, ESA,\\ https://xmm-tools.cosmos.esa.int/external/\\xmm\_user\_support/documentation/uhb/

\bibitem[Yakovlev \& Pethick(2004)]{Yakovlev04} Yakovlev, D.~G., \& Pethick, C.~J.\ 2004, \araa, 42, 169

\bibitem[Young(1972)]{Young72} Young, M.\ 1972, Journal of the Optical Society of America (1917-1983), 62, 972 

\end{thebibliography}
\end{document}